\documentclass[journal,12pt,draftclsnofoot,onecolumn,english]{IEEEtran}
\usepackage{graphicx}
\usepackage{amsmath,amssymb}
\allowdisplaybreaks[1]
\usepackage{cases}
\usepackage{color}
\usepackage{amsfonts}
\usepackage{indentfirst}
\usepackage{cite}
\usepackage{caption}
\usepackage{algorithm}
\usepackage{algpseudocode}
\usepackage{booktabs}
\usepackage{blkarray}
\usepackage{subfigure}
\usepackage{multirow}
\usepackage{amsthm}
\usepackage{amssymb}
\usepackage{enumerate}
\usepackage{mathrsfs}
\usepackage{pifont}
\usepackage{multirow}
\usepackage{booktabs}
\usepackage{bigstrut}
\usepackage{hyperref}
\usepackage[table]{xcolor}

\makeatletter
\newtheorem{theorem}{Theorem}
\newtheorem{example}{Example}
\newtheorem{definition}{Definition}
\newtheorem{remark}{Remark}

\newtheorem{lemma}{Lemma}

\begin{document}
	\title{Coded Caching Schemes for Two-dimensional Caching-aided Ultra-Dense Networks
		\author{Minquan Cheng, Jinwei Xu, Mingming Zhang, and Youlong Wu}
		\thanks{M. Cheng, J. Xu and M. Zhang are with Guangxi Key Lab of Multi-source Information Mining $\&$ Security, Guangxi Normal University,
		Guilin 541004, China (e-mail:  chengqinshi@hotmail.com, mjml\_xujinwei@163.com, ztw\_07@foxmail.com).		}
		\thanks{Youlong Wu is with the School of Information Science and Technology, ShanghaiTech University,
				201210 Shanghai, China. (e-mail: wuyl1@shanghaitech.edu.cn).}
	}
	\date{}
	\maketitle

\begin{abstract}
Coded caching technique is an efficient approach to reduce the transmission load in networks and has been studied in heterogeneous network settings in recent years. In this paper, we consider a new widespread caching system called $(K_1,K_2,U,r,M,N)$ two-dimensional (2D) caching-aided ultra-dense network (UDN) with a server containing $N$ files, $K_1K_2$ cache nodes arranged neatly on a grid with $K_1$ rows and $K_2$ columns, and $U$ cache-less users randomly distributed around cache nodes. Each cache node can cache at most $M\leq N$ files and has a certain service region by Euclidean distance. The server connects to users through an error-free shared link and the users in the service region of a cache node can freely retrieve all cached contents of this cache node. We aim to design a coded caching scheme for 2D caching-aided UDN systems to reduce the transmission load in the worst case while meeting all possible users' demands. First, we divide all possible users into four classes according to their geographical locations. Then our first order optimal scheme is proposed based on the Maddah-Ali and Niesen scheme. Furthermore, by compressing the transmitted signals of our first scheme based on Maximum Distance Separable (MDS) code, we obtain an improved order optimal scheme with a smaller transmission load.
\end{abstract}

\section{Introduction}
Driven by the dramatic increase in intelligent devices and fast development of application services such as autonomous driving, ultra-high-definition video broadcasting, and the Internet of things, mobile data traffic witnesses unprecedented growth which will saturate the network capacity. To address this challenge, caching is regarded as an effective approach to reduce peak traffic time by prefetching popular content into memories during the off-traffic time.

Recently, coded caching scheme has been proposed by Maddah-Ali and Niesen (referred as to the MN scheme) for the $(K,M,N)$ cache-aided broadcast network where a single-antenna server has a library of $N$ equal-sized files and broadcasts to $K$ users each of cache size $M$ files through an error-free shared link in \cite{MN}. The MN scheme applies coding techniques to create multicast opportunities to further reduce the transmission pressure during the peak-traffic time.  A coded caching scheme contains a placement phase during the off-peak time and a delivery phase during the peak-traffic time. In the placement phase, the server places some contents to each user's cache without knowledge of the users' future demands. In the delivery phase, each user requests one arbitrary file and the server broadcasts coded signals such that each user can recover its desired file. The worst-case load over all possible demands is defined as the {\em transmission load}, that is, the number of files that must be communicated so that any demands can be satisfied. It was proved in \cite{yufactor2TIT2018} that the MN scheme achieves the optimal transmission load within a multiplicative gap of at most $2$ when $N\geq K$.  In addition, under the constraint of uncoded cache placement (i.e., each user directly caches a subset of the library bits) the MN scheme was proved to be optimal in \cite{WTP,YMA} when $N\geq K$. So the MN scheme has been widely used in different topological networks such as device-to-device networks in \cite{JCM}, combination networks in \cite{JAJC,WJPD},  hierarchical networks in \cite{NUMS}, etc.

Ultra-dense network (UDN) is another key technology to cope with the unprecedented growth of mobile data traffic, and is widely applied in modern cellular systems \cite{UDN'12,KHY}. By densely deploying multiple access points (e.g., pico-cell base stations and femto-cell access points) to offload data traffic of the macro base stations (MBSs), UDN can significantly improve the system throughput per area. Recent works have proposed to equip access points with cache spaces (referred as to cache nodes), and applied coded caching techniques into UDN to improve the spectral efficiency and reduce communication latency \cite{MV1979,JHNS,SPE,RK,SR,RKstructure,OG,CWLZC,NRsecure,KMR,MKR,MKRmn,NRprivacy,FP,WCLC,ZWCC,WCKC}. In \cite{JHNS}, the authors first proposed a $(K,L,M,N)$ one-dimensional multi-access coded caching (1D MACC) scheme for UDN where there are $K$ cache nodes each of a cache size of $M$ files, while each of the $K$ users is cache-less and can access $L$ neighboring cache nodes in a cyclic wrap-around fashion. Some other works considered different accessing topologies for UDN. For example, \cite{MKRmn} investigated a combination accessing topology where there are exactly ${K\choose L}$ users, each of which accesses a unique cache node set of size $L$.  \cite{FP} considered all the possible accessing topologies where there are exactly $\sum_{h=0}^{L}{K\choose h}$ users, each of which accesses a unique cache node set of size $0\leq h\leq L$. There are some works extending the 1D MACC to two-dimensional (2D) MACC \cite{MV1979,ZWCC}. In particular, the authors in \cite{ZWCC} considered $(K_1,K_2,L,M,N)$ 2D caching-aided UDN where $K_1\times K_2$ cache nodes with a cache size of $M$ files are placed in a rectangular grid with $K_1$ rows and $K_2$ columns, and $K_1\times K_2$ cache-less users are placed regularly on the same grid such that each user is in the proximity of a square of $L\times L$ neighboring cache nodes (the distance is defined in a cyclic wrap-around fashion). It is worth noting that all the above works assume that each user is served by the same number of cache nodes, ignoring the fact that the number could vary depending on users' geographical locations. To address this issue, we propose a coded caching scheme for more practical 2D caching-aided UDN system where the users can associate with various numbers of cache nodes according to the facility locations.

In this paper, we focus on a $(K_1, K_2, U, r,M, N)$ 2D caching-aided UDN (see Fig. \ref{fig:model}), which includes a server (e.g., MBS) containing $N$ files of equal size, $K_1K_2$ cache nodes arranged in a grid of $K_1$ rows and $K_2$ columns each of which caches $M$ files where $0 \leq M \leq N$ (same as in  \cite{ZWCC}), and $U$ mobile devices each of which is randomly distributed around cache nodes (different from the prior works). We assume that the server connects to $U$ users through an error-free shared link, the distance between any two consecutive cache nodes is a unit in horizontal and vertical directions, the service radius of each cache node is the real number $r$ where $\frac{\sqrt{2}}{2}\leq r \leq 1$, and user devices access the cache node within modular Euclidean distance (see Definition \ref{DefMED}) $r$ in cyclic wrap-around topology $K_1$ and $K_2$ respectively.  Similar to prior works \cite{MV1979,KMR,MKR,MKRmn,FP,ZWCC} that ignored the transmission between the cache nodes and users, our goal is to reduce the transmission load between the server and cache nodes. The main contributions are summarized as follows.

\begin{itemize}
	\item According to the service radius of each cache node, we show that all the possible users can be divided into four classes. Based on these classes, we design a coded caching scheme, referred as to Scheme A, for $(K_1, K_2, U, r,M, N)$ 2D caching-aided UDN with $U=3K_1K_2$, $8K_1K_2$ and $7K_1K_2$ if $r=\frac{\sqrt{2}}{2}$, $\frac{\sqrt{2}}{2}< r < 1$ and $r=1$ respectively,  each of which achieves the optimal transmission load within a multiplicative gap of at most $48$, i.e., Scheme A is order optimal. In addition, it is worth noting that Scheme A can be used to the case for any number of users.
	\item Since some users could access multiple cache nodes, their contents in the transmitted signals can be retrieved by these users. In order to reduce some redundancy distributed in the transmitted signals of Scheme A,  we proposed a coded caching scheme,  namely Scheme B, based on the  Maximum Distance Separable (MDS) code, which can obtain a smaller transmission load compared to Scheme A.
\end{itemize}

The rest of this paper is organized as follows. Section \ref{sec-Shared-link-system} reviews the original shared-link coded caching model and our proposed $2$D caching-aided UDN system. Section  \ref{sec-baseline} introduces Scheme A for the proposed system and its theoretic performance analysis. Section \ref{sec-Improved} proposes a Scheme B based on Scheme A. Finally, the conclusion is given in Section \ref{sec-conlusion}.

First the following notations are used in this paper unless otherwise stated.
Bold lower case letter and curlicue font will be used to denote vector and set respectively.
$|\cdot|$ is used to represent the cardinality of set or the length of vector.
For any positive integers $a,b$ with $a<b$, and non-negative set $\mathcal{K}$,
$[a:b)=\{a,a+1,\ldots,b-1\}$, $[a:b]=\{a,a+1,\ldots,b-1,b\}$,
${\mathcal{K} \choose t}=\{\mathcal{V}\subseteq \mathcal{K}\ |\ |\mathcal{V}|=t\}$, i.e., ${\mathcal{K} \choose t}$ is the collection of all t-sized subsets of $\mathcal{K}$.
$<a>_q=\text{mod}(a,q)$ denotes the least non-negative residue of $a$ modulo $q$.

\section{Shared-link caching system and $2$D caching-aided UDN system}
\label{sec-Shared-link-system}
In this section, we introduce the original shared-link caching system and its related MN coded caching scheme, and $2$D caching-aided UDN system respectively.

\subsection{Shared-link Caching System and MN Scheme}
In a $(K,M,N)$ shared-link coded caching system~\cite{MN}, a server containing $N$ files $\mathcal{W}=\{W_{0}, W_{1}, \ldots, $ $W_{N-1}\}$ with equal length connects through an error-free shared link to $K$ users $\{\text{U}_0,\text{U}_1,\ldots,\text{U}_{K-1}\}$ where $K\leq N$, and each user has a cache device which can store up $M$ files with $0\leq M \leq N$. An $F$-division $(K,M,N)$ coded caching scheme contains two phases.
\begin{itemize}
	\item Placement phase: Each file is divided into $F$ packets with equal size, and then each user $\text{U}_k$ where $k\in [0:K)$ caches some packets of each file, which is limited by its cache size $M$\footnote{\label{foot:uncoded}In this paper, we only consider the uncoded cache placement.}. Let $\mathcal{Z}_{\text{U}_k}$ denote the cache contents at user $\text{U}_k$, which is assumed to be known to the server. Notice that the placement phase is done without knowledge of later requests.
	\item Delivery phase: Each user randomly requests a file from the server. The requested file by user $\text{U}_k$ is represented by $W_{d_{\text{U}_k}}$, and the request vector by all users is denoted by $\mathbf{d}=(d_{\text{U}_0},d_{\text{U}_1},\ldots,d_{\text{U}_{K-1}})$. According to the cached contents and request vector, the server broadcasts $S_{{\bf d}}$ coded packets to all users such that each user's request is satisfied.
\end{itemize}
In such system, the number of worst-case transmitted files (a.k.a. transmission load) for all possible requests is expected to be as small as possible, which is defined as
\begin{align}
	R=\max_{\mathbf{d}\in[0:N)^K} \frac{ S_{\mathbf{d}}}{F} . \label{eq:def of load}
\end{align}
In \cite{MN}, Maddah-Ali and Niesen proposed a caching scheme for the $(K,M,N)$ coded caching system under the assumption that the files can be arbitrarily subdivided. Algorithm \ref{alg-MN} depicts the ${K\choose KM/N}$-MN scheme where $M/N\in\{0, 1/K,2/K,\cdots,1\}$.
It was shown in \cite{MN} that, the scheme is feasible and each user can successfully recover its requested file at a load
\begin{eqnarray}
\label{eq:MN}
R_{\text{MN}}=\frac{K(1-\frac{M}{N})}{\frac{KM}{N}+1}=\frac{K-t}{t+1}, \ \forall M=\frac{N t}{K}: t\in [0:K].
\end{eqnarray} For general $0\leq M/N\leq 1$, the lower convex envelop of the following memory-load tradeoff corner point is achievable by the MN scheme.

\begin{algorithm}[htb]
	\caption{The MN Coded Caching Scheme \cite{MN}}\label{alg-MN}
	\begin{algorithmic}[1]
		\State $t\leftarrow\frac{KM}{N}$
		\Procedure {Placement}{$W_0,\cdots,W_{N-1}$}
		\For{$n\in[0,N)$}
		\State Split $W_n$ into $\{W_{n,\mathcal{T}}\ |\ \mathcal{T}\in{[0:K)\choose t}\}$ of equal packets
		\EndFor
		\For{$k\in[0:K)$}
		\State $\mathcal{C}_k\leftarrow\{W_{n,\mathcal{T}}\ |\ n\in[0,N),\mathcal{T}\in{[0:K)\choose t},k\in\mathcal{T}\}$
		\EndFor
		\EndProcedure
		\Procedure{Delivery}{$W_0,\cdots,W_{N-1},d_{\text{U}_0},\cdots,d_{\text{U}_{K-1}}$}
		\State Server sends $\{X_{\mathcal{S}}=\oplus_{k\in\mathcal{S}} W_{d_{\text{U}_k},\mathcal{S}\backslash \{k\}}\ |\ \mathcal{S}\in{[0:K)\choose t+1}\}$
		\EndProcedure
	\end{algorithmic}
\end{algorithm}

For clarity, we give an example from \cite{MN} to illustrate the scheme.

\begin{example}\rm \label{exam-MN}
	When $N=K=3$ and $M=1$, there are three files $W_0$, $W_1$, $W_2$ and three users $\text{U}_0$, $\text{U}_1$, $\text{U}_2$ which the cache size is $M=1$ file. By Algorithm \ref{alg-MN}, we have $t=\frac{KM}{N}=1$ and obtain  following $(K,M,N)=(3,1,3)$ MN scheme.
\begin{itemize}
\item Placement phase: Each file is divided into $F={K\choose t}=3$ packets with equal size, i.e., $W_0=\{W_{0,\{0\}},W_{0,\{1\}},W_{0,\{2\}}\}$, $W_1=\{W_{1,\{0\}},W_{1,\{1\}},W_{1,\{2\}}\}$ and $W_2=\{W_{2,\{0\}},W_{2,\{1\}},W_{2,\{2\}}\}$. Then by Line 7 of Algorithm \ref{alg-MN}, the users cache $\mathcal{Z}_{\text{U}_0}=$ $\{W_{0,\{0\}}$, $W_{1,\{0\}}$, $W_{2,\{0\}}\}$, $\mathcal{Z}_{\text{U}_1}=$ $\{W_{0,\{1\}}$, $W_{1,\{1\}}$, $W_{2,\{1\}}\}$ and $\mathcal{Z}_{\text{U}_2}=$ $\{W_{0,\{2\}}$, $W_{1,\{2\}}$, $W_{2,\{2\}}\}$ respectively. Clearly the cache content of each user is exactly one file.
\item Delivery phase: Assume that the request vector $\mathbf{d}=(0,1,2)$. Then the server sends the following $S={K\choose t+1}=3$ coded packets for all users,
	$$
	X_{\{0,1\}}=W_{0,\{1\}}\bigoplus W_{1,\{0\}},\ \ X_{\{0,2\}}=W_{0,\{2\}}\bigoplus W_{2,\{0\}},\ \ X_{\{1,2\}}=W_{1,\{2\}}\bigoplus W_{2,\{1\}}.
	$$
	The user $\text{U}_0$ can recover the requested file $W_0$ according to its own cached content $\mathcal{Z}_{\text{U}_0}$ and the two broadcast messages $X_{\{0,1\}}$, $X_{\{0,2\}}$. Similarly, the users $\text{U}_1$ and $\text{U}_2$ can also recover the requested files according to $\mathcal{Z}_{\text{U}_1}$, $X_{\{0,1\}}$, $X_{\{1,2\}}$ and $\mathcal{Z}_{\text{U}_2}$, $X_{\{0,2\}}$, $X_{\{1,2\}}$ respectively. So we can get the transmission load of this scheme is $R=\frac{S}{F}=\frac{3}{3}=1$.
	\end{itemize}

\end{example}
\subsection{The 2D Cache-aided UDN System}

\begin{figure}[h]
	\centering
	\includegraphics[scale=0.25]{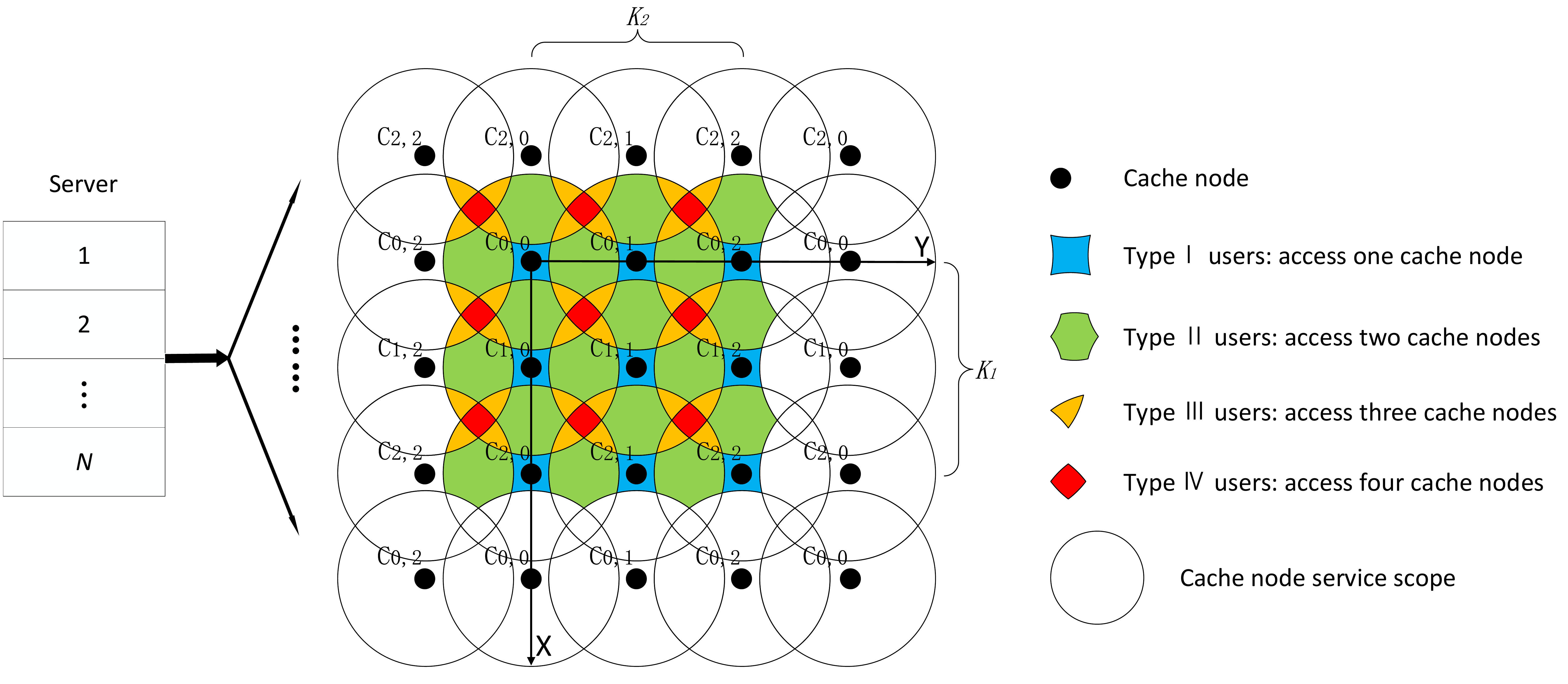}
	\caption{The $(K_1, K_2, U,r, M, N)$ 2D caching-aided UDN system with $\frac{\sqrt{2}}{2}< r < 1$ and $K_1=K_2=3$.}
	\label{fig:model}
\end{figure}

In a $(K_1, K_2, U,r, M, N)$ 2D caching-aided UDN system in Fig. \ref{fig:model},
 there is a single server containing $N$ files with equal size, denoted as $\mathcal{W} = \{W_0, \dots,W_{N-1}\}$, $K_1K_2$ cache nodes $\text{C}_{k_1,k_2}$ where $(k_1,k_2)\in \mathcal{K}=[0:K_1)\times [0:K_2)$ and $K_1,K_2\geq 3$ orderly arranged in a rectangular array of $K_1$ rows and $K_2$ columns, each of which caches $M$ files where $0 \leq M \leq N$, and $U$ users where $U\leq N$ randomly distributed around cache nodes. We assume that the server connects to $K_1K_2$ cache nodes through an error-free shared link, the distance between any two consecutive cache nodes is one unit in the horizontal and vertical directions. Then the distance between two consecutive cache nodes is $\sqrt{2}$ in a diagonal direction. In order to make that any located user can be served at least one cache node, we assume that the service radius of each cache node is $r$ where $\frac{\sqrt{2}}{2}\leq r \leq 1$. Then users access the cache node within modular Euclidean distance $r$ under $K_1$ and $K_2$ in cyclic wrap-around topology, where the modular Euclidean distance is defined as follows.

\begin{definition}[{\em Modular Euclidean Distance}]\label{DefMED}\rm
For any integers $0\leq k_1,k'_1<K_1$ and $0\leq k_2,k'_2<K_2$, the modular Euclidean distance under $K_1$ and $K_2$ respectively is defined as $$\text{d}_{K_1,K_2}((k_1,k_2),(k'_1,k'_2))=\sqrt{(\text{d}_{K_1}(k_1,k'_1))^2
 +(\text{d}_{K_2}(k_2,k'_2))^2}$$
Here for any integers $0\leq k\leq k'<K$, $\text{d}_{K}(k,k')=\min\{k'-k,k-k'+K\}$.
\end{definition}

For any possible subset $\mathcal{C}\subseteq \mathcal{K}$, let $\lambda_{\mathcal{C}}$ represents the number of users each of which can access all the cache nodes labeled by $\mathcal{C}$. In this paper we focus on designing the scheme for a $(K_1, K_2, U,r, M, N)$ 2D caching-aided UDN system with $\lambda_{\mathcal{C}}=1$ for all the possible subset $\mathcal{C}\subseteq\mathcal{K}$. For the case $|\lambda_{\mathcal{C}}|> 1$, there exist multiple users accessing the same set of cache nodes, i.e., receiving the same signals, these two users can not generate any multicast opportunities. Hence we can use the same delivery strategies $\lambda$ times by adding some virtual users where $\lambda$ is the maximum value among the values of $\lambda_{\mathcal{C}}$ for all the possible subset $\mathcal{C}\subseteq \mathcal{K}$. So it is sufficient to design the scheme for the case $\lambda_{\mathcal{C}}=1$ for all the possible subset $\mathcal{C}\subseteq\mathcal{K}$. In this way, we use the index set $\{(k_1,k_2) | (k_1,k_2)\in\mathcal{C}\}$ to represent the user served exclusively by the servers in cache nodes labeled by $\mathcal{C}$.

\begin{lemma}\rm
	\label{lemma:r-U}
	In this coded caching system, we only need to design the transmission scheme for the following number of users.
	\begin{align*}
		U=
		\left\{
		\begin{array}{lr}
			3K_1K_2, \qquad \text{if }\ r=\frac{\sqrt{2}}{2},\\
			8K_1K_2, \qquad \text{if }\ \frac{\sqrt{2}}{2}< r < 1,\\
			7K_1K_2, \qquad \text{if }\ r=1.
		\end{array}
		\right.
	\end{align*}
\end{lemma} The detailed explanation can be found in \ref{sec-user's types}.

An $F$-division $(K_1, K_2, U,r, M, N)$ 2D caching-aided UDN scheme consists of the following two phases.
\begin{itemize}
\item Placement phase: Each file is divided into $F$ packets with equal size, each cache node $\text{C}_{k_1,k_2}$ where $(k_1,k_2)\in \mathcal{K}$ directly caches some packets of each file, which is limited by its cache size $M$ and the cached contents in cache node denoted by $\mathcal{Z}_{\text{C}_{k_1,k_2}}$. Each user retrieves the cached contents from one or more connected cache nodes. In this phase, the server has no knowledge of later users' requests.
\item Delivery phase: Each user randomly requests a file. According to the users' requests and the contents retrieved by users, the server transmits coded packets with the size of at most $R$ files to users such that each user can recover the requested file.
\end{itemize}

\begin{definition}[Optimality load]\rm
The pair $(\mathcal{M}, R)$ is achievable, for any real number $\epsilon > 0$ there exists a caching scheme such that every user is able to reconstruct its requested file with a probability of error less than $\epsilon$. The load tradeoff is defined as
\begin{align*}
R^*(K,\mathcal{M},N)\triangleq \inf \{R\ |\ (\mathcal{M},R)\ \text{is achievable}\}.
\end{align*}
\end{definition}

We aim to design a coded caching scheme with transmission load defined in~\eqref{eq:def of load} as small as possible.

\section{Scheme A for 2D caching-aided UDN system}
\label{sec-baseline}
In this section
according to the service region, we first show that all the possible users of a $(K_1, K_2, U,r, M, N)$ 2D caching-aided UDN system can be divided into four classes. Then based on these four classes, we propose our first order optimal coded caching scheme, i.e., Scheme A. Finally, some numerical performances are proposed.
\subsection{The Classification of User}
\label{sec-user's types}
We consider a 2D caching-aided UDN system under Cartesian coordinate system, where each $(k_1,k_2)\in\mathcal{K}$ is presented as the cache node in the $k_1$th row and $k_2$th column. For instance, when $K_1=K_2=3$, the vector $(0,0)$ represents the cache node $\text{C}_{0,0}$ in Fig. \ref{fig:model}. Then each user located in the service region of cache node $\text{C}_{k_1,k_2}$ can be denoted by $\text{U}_{k_1+x,k_2+y}$ where $x,y\in \mathbb{R}$ satisfy $\sqrt{x^2+y^2}\leq r$. In addition, the user $\text{U}_{k_1+x,k_2+y}$ is also in the service region of another cached node $\text{C}_{k'_1,k'_2}$ if and only if the following inequality holds,
\begin{align}\label{eq-serve-condition}
\text{d}_{K_1,K_2}((k_1+x,k_2+y),(k'_1,k'_2))=
\sqrt{(\text{d}_{K_1}(k_1+x,k'_1))^2+(\text{d}_{K_2}(k_2+y,k'_2))^2}
\leq r.
\end{align}Recall that for any two integers $k$, $k'$ and any positive $K$ with $0\leq k\leq k'<K$, their modular Euclidean distance under $K$, $\text{d}_{K}(k,k')=\min\{k'-k,k-k'+K\}$.

When $\frac{\sqrt{2}}{2}< r < 1$, each user $\text{U}_{k_1+x,k_2+y}$ where $x,y\in \mathbb{R}$ satisfying $\sqrt{x^2+y^2}< r$ and $(k_1,k_2)\in \mathcal{K}$ can access some cache nodes from the nine possible cache nodes
\begin{align*}
\begin{array}{ccc}
\text{C}_{<k_1-1>_{K_1},<k_2-1>_{K_2}},& \text{C}_{<k_1-1>_{K_1},k_2},& \text{C}_{<k_1-1>_{K_1},<k_2+1>_{K_2}},\\
\text{C}_{k_1,<k_2-1>_{K_2}},\ \ \ \ \ \ \ \ ~ & \text{C}_{k_1,k_2},\ \ \ \ \ \ \ \ ~ & \text{C}_{k_1,<k_2+1>_{K_2}},\ \ \ \ \ \ \ \ ~ \\
\text{C}_{<k_1+1>_{K_1},<k_2-1>_{K_2}},& \text{C}_{<k_1+1>_{K_1},k_2},& \text{C}_{<k_1+1>_{K_1},<k_2+1>_{K_2}}.
\end{array}
\end{align*}
From \eqref{eq-serve-condition}, we define the distances from user $\text{U}_{k_1+x,k_2+y}$ to the above nine cache nodes as follows.
\begin{align}
	\label{eq-righ-up-down-all}
	\begin{split}
e_{-1,-1}(x,y)&\triangleq\text{d}_{K_1,K_2}((k_1+x,k_2+y),(k_1-1,k_2-1))=\sqrt{(1+x)^2+(1+y)^2},\\
e_{-1,0}(x,y)&\triangleq\text{d}_{K_1,K_2}((k_1+x,k_2+y),(k_1-1,k_2))\ \ \ \ \ =\sqrt{(1+x)^2+y^2},\\
e_{-1,1}(x,y)&\triangleq\text{d}_{K_1,K_2}((k_1+x,k_2+y),(k_1-1,k_2+1)) =\sqrt{(1+x)^2+(1-y)^2},\\
e_{0,-1}(x,y)&\triangleq\text{d}_{K_1,K_2}((k_1+x,k_2+y),(k_1,k_2-1))\ \ \ \ \ =\sqrt{x^2+(1+y)^2},\\
e_{0,0}(x,y)&\triangleq\text{d}_{K_1,K_2}((k_1+x,k_2+y),(k_1,k_2))\qquad \ \ \ \ =\sqrt{x^2+y^2},\\
e_{0,1}(x,y)&\triangleq\text{d}_{K_1,K_2}((k_1+x,k_2+y),(k_1,k_2+1))\ \ \ \ \ =\sqrt{x^2+(1-y)^2},\\
e_{1,-1}(x,y)&\triangleq\text{d}_{K_1,K_2}((k_1+x,k_2+y),(k_1+1,k_2-1))=\sqrt{(1-x)^2+(1+y)^2},\\
e_{1,0}(x,y)&\triangleq\text{d}_{K_1,K_2}((k_1+x,k_2+y),(k_1+1,k_2))\quad \ \ =\sqrt{(1-x)^2+y^2},\\
e_{1,1}(x,y)&\triangleq\text{d}_{K_1,K_2}((k_1+x,k_2+y),(k_1+1,k_2+1)) =\sqrt{(1-x)^2+(1-y)^2}.
\end{split}
\end{align}
By checking the values in \eqref{eq-righ-up-down-all} each of which is less than or equal to $r$,
we can divide all the possible users in a $(K_1,K_2,U,r,M,N)$ 2D caching-aided UDN system into the following four types. For detailed proof, please see Appendix \ref{appendix-user-type}.
\begin{itemize}
\item {\bf Type I:} Each user can access exactly one cache node. Tn this case, all the users can be represented by
\begin{align}\label{eq-type-I}
\mathcal{L}_{\text{I}}=\left\{\mathcal{U}^{\text{I}}_{k_1,k_2}=\{(k_1,k_2)\}\ |\ (k_1,k_2)\in \mathcal{K}\right\}.
\end{align}
\item {\bf Type II:} Each user can access exactly two different cache nodes. In addition, all the users of Type II can be divided into two sub-types and represented by
\begin{subequations}
\label{eq-type-II}
\begin{align}
\label{eq-type-II-1}
\mathcal{L}_{\text{II-1}}=\left\{\mathcal{U}^{\text{II}-1}_{k_1,k_2}=\{(k_1,k_2),(k_1,<k_2+1>_{K_2})\}\ |\ (k_1,k_2)\in \mathcal{K}\right\},\\
\label{eq-type-II-2}
\mathcal{L}_{\text{II-2}}=\left\{\mathcal{U}^{\text{II}-2}_{k_1,k_2}=\left\{(k_1,k_2),(<k_1+1>_{K_1},k_2)\right\}\ |\ (k_1,k_2)\in \mathcal{K}\right\}.
\end{align}
\end{subequations}

\item {\bf Type III:} Each user can access exactly three different cache nodes. In addition, all the users of Type III can be divided into the four sub-types and represented by
\begin{subequations}
\label{eq-type-III}
\begin{align}
\mathcal{L}_{\text{III-1}}&=\left\{\mathcal{U}^{\text{III}-1}_{k_1,k_2}=\left\{(k_1,k_2),
(k_1,<k_2+1>_{K_2}),(<k_1+1>_{K_1},k_2)\right\} | (k_1,k_2)\in \mathcal{K}\right\} \label{eq-type-III-1},\\
\mathcal{L}_{\text{III-2}}&=\left\{\mathcal{U}^{\text{III}-2}_{k_1,k_2}=\left\{(k_1,k_2),(<k_1+1>_{K_1},k_2),(<k_1+1>_{K_1},<k_2+1>_{K_2})\right\} | (k_1,k_2)\in \mathcal{K}\right\}\label{eq-type-III-2},\\
\mathcal{L}_{\text{III-3}}&=\left\{\mathcal{U}^{\text{III}-3}_{k_1,k_2}=\left\{(k_1,k_2),(k_1,<k_2+1>_{K_2}),(<k_1+1>_{K_1},<k_2+1>_{K_2})\right\} | (k_1,k_2)\in \mathcal{K}\right\}\label{eq-type-III-3},\\
\mathcal{L}_{\text{III-4}}&=\left\{\mathcal{U}^{\text{III}-4}_{k_1,k_2}=\left\{(k_1,k_2),(k_1,<k_2-1>_{K_2}),(<k_1-1>_{K_1},k_2)\right\}\ |\ (k_1,k_2)\in \mathcal{K}\right\}.
\label{eq-type-III-4}
\end{align}
\end{subequations}
\item {\bf Type IV:} Each user can access four different cache nodes. In addition, all the users of Type IV can be represented by
\begin{align}\label{eq-type-IV}	
&\mathcal{L}_{\text{IV}}=\left\{\mathcal{U}^{\text{IV}}_{k_1,k_2}=\left\{(k_1,k_2),
(<k_1+1>_{K_1},k_2),(k_1,<k_2+1>_{K_2}),(<k_1+1>_{K_1},\right.\right. \nonumber \\
&\ \ \ \ \ \ \ \ \left.\left.<k_2+1>_{K_2}>)\right\}\ |\ (k_1,k_2)\in \mathcal{K}\right\}.
\end{align}
\end{itemize}

From the discussion in  Appendix \ref{appendix-user-type}, when $\frac{\sqrt{2}}{2}<r<1$, the set of user in a $(K_1,K_2,U,M,N)$ 2D caching-aided UDN system is
\begin{align}
\label{allusers}
\mathcal{L}=\mathcal{L}_{\text{I}}\bigcup\mathcal{L}_{\text{II-1}}\bigcup \mathcal{L}_{\text{II-2}}\bigcup \mathcal{L}_{\text{III-1}}\bigcup\mathcal{L}_{\text{III-2}}\bigcup\mathcal{L}_{\text{III-3}}
\bigcup\mathcal{L}_{\text{III-4}}\bigcup\mathcal{L}_{\text{IV}}.
\end{align}
That is, there are $K_1K_2$ users of Type I, $2K_1K_2$ users of Type II, $4K_1K_2$ users of Type III and $K_1K_2$ users of Type IV. Then the total number of users $U=8K_1K_2$ in $(K_1,K_2,U,r,M,N)$ 2D caching-aided UDN system when $\frac{\sqrt{2}}{2}< r < 1$.

By Appendix \ref{appendix-user-type}, the cases for $r=\frac{\sqrt{2}}{2}$ and $1$ can be directly obtained as follows.
\begin{remark}\rm
\label{usertype_r=limit}
We introduce the two special cases of each cache node radius $r=\frac{\sqrt{2}}{2}$ and $r=1$ in $(K_1,K_2,U,r,M,N)$ 2D caching-aided UDN system.
\begin{itemize}
\item When cache node radius $r=\frac{\sqrt{2}}{2}$, by Appendix \ref{appendix-user-type} the equations have no solution in \eqref{eq-tyep-III-codition-1}, \eqref{eq-tyep-III-codition-2}, \eqref{eq-tyep-III-codition-3}, \eqref{eq-tyep-III-codition-4} and \eqref{eq-tyep-IV-codition-1}, then there are no users access $3$ or $4$ cache nodes, i.e., the users of Type III and Type IV do not exist in the system model. Thus, the user set is
\begin{align}
\label{allusers_min}
\mathcal{L}=\mathcal{L}_{\text{I}}\bigcup\mathcal{L}_{\text{II-1}}\bigcup \mathcal{L}_{\text{II-2}}.
\end{align}
which contains $K_1K_2$ Type I and $2K_1K_2$ Type II users, i.e., $U=3K_1K_2$.

\item When cache node radius $r=1$, and Appendix \ref{appendix-user-type}, the equations have no solution in \eqref{eq-tyep-I-codidtion-1}, then there is no user accessing exactly one cache node, i.e., the users of Type I do not exist in the system model. Therefore, the user set is
\begin{align}
\label{allusers_max}
\mathcal{L}=\mathcal{L}_{\text{II-1}}\bigcup\mathcal{L}_{\text{II-2}}\bigcup \mathcal{L}_{\text{III-1}}\bigcup\mathcal{L}_{\text{III-2}}\bigcup
\mathcal{L}_{\text{III-3}}\bigcup\mathcal{L}_{\text{III-4}}\bigcup\mathcal{L}_{\text{IV}}.
\end{align}  That is, there are $2K_1K_2$ Type II users, $4K_1K_2$ Type III users and $K_1K_2$ Type IV users, i.e., $U=7K_1K_2$.
\end{itemize}
\end{remark}

\begin{example}\rm\label{e.g-users}
Let us see the $(K_1, K_2, U,r, M, N)=(3,3,72,r,16,72)$ 2D caching-aided UDN system with $\frac{\sqrt{2}}{2}< r < 1$ listed in Fig. \ref{fig:model}. From \eqref{eq-type-I}, \eqref{eq-type-II}, \eqref{eq-type-III} and \eqref{eq-type-IV}, all users can be obtained as follows.

\noindent $\bullet$ From \eqref{eq-type-I} the user set of Type I in the blue regions of Fig. \ref{fig:model} is
\begin{align*}
\mathcal{L}_{\text{I}}=&\left\{\mathcal{U}^{\text{I}}_{0,0}=\{0,0\},\  \mathcal{U}^{\text{I}}_{0,1}=\{0,1\},\ \mathcal{U}^{\text{I}}_{0,2}=\{0,2\},\
\mathcal{U}^{\text{I}}_{1,0}=\{1,0\},\ \mathcal{U}^{\text{I}}_{1,1}=\{1,1\},\right.\\
&\ \left.
\mathcal{U}^{\text{I}}_{1,2}=\{1,2\},\ \mathcal{U}^{\text{I}}_{2,0}=\{2,0\},\ \mathcal{U}^{\text{I}}_{2,1}=\{2,1\},\ \mathcal{U}^{\text{I}}_{2,2}=\{2,2\}\right\}.
\end{align*}

\noindent $\bullet$ From \eqref{eq-type-II} the user sets of Type II in the green regions of Fig. \ref{fig:model} are
\begin{align*}
\text{Type II-1:}\ \mathcal{L}_{\text{II-1}}=&\left\{
\mathcal{U}^{\text{II-1}}_{0,0}=\{(0,0),(0,1)\},\ \mathcal{U}^{\text{II-1}}_{0,1}=\{(0,1),(0,2)\},\ \mathcal{U}^{\text{II-1}}_{0,2}=\{(0,2),(0,0)\},\right.\\
&\ \ \mathcal{U}^{\text{II-1}}_{1,0}=\{(1,0),(1,1)\},\ \mathcal{U}^{\text{II-1}}_{1,1}=\{(1,1),(1,2)\},\ \mathcal{U}^{\text{II-1}}_{1,2}=\{(1,2),(1,0)\},\\
&\ \left.\mathcal{U}^{\text{II-1}}_{2,0}=\{(2,0),(2,1)\},\ \mathcal{U}^{\text{II-1}}_{2,1}=\{(2,1),(2,2)\},\ \mathcal{U}^{\text{II-1}}_{2,2}=\{(2,2),(2,0)\}\right\};
\end{align*}
\begin{align*}
\text{Type II-2:}\ \mathcal{L}_{\text{II-1}}=&\left\{
\mathcal{U}^{\text{II-1}}_{0,0}=\{(0,0),(1,0)\},\ \mathcal{U}^{\text{II-1}}_{0,1}=\{(0,1),(1,1)\},\ \mathcal{U}^{\text{II-1}}_{0,2}=\{(0,2),(1,2)\},\right.\\
&\ \ \mathcal{U}^{\text{II-1}}_{1,0}=\{(1,0),(2,0)\},\ \mathcal{U}^{\text{II-1}}_{1,1}=\{(1,1),(2,1)\},\ \mathcal{U}^{\text{II-1}}_{1,2}=\{(1,2),(2,2)\},\\
&\ \left.\mathcal{U}^{\text{II-1}}_{2,0}=\{(2,0),(0,0)\},\ \mathcal{U}^{\text{II-1}}_{2,1}=\{(2,1),(0,1)\},\ \mathcal{U}^{\text{II-1}}_{2,2}=\{(2,2),(0,2)\}\right\}.
\end{align*}

\noindent $\bullet$ From \eqref{eq-type-III} the user sets of Type III in the yellow regions of Fig. \ref{fig:model} are
\begin{align*}
\text{Type III-1:}\ \mathcal{L}_{\text{III-1}}=&\left\{
\mathcal{U}^{\text{III-1}}_{0,0}=\{(0,0),(0,1),(1,0)\},\ \mathcal{U}^{\text{III-1}}_{0,1}=\{(0,1),(0,2),(1,1)\},\right.\\
&\ \left.\mathcal{U}^{\text{III-1}}_{0,2}=\{(0,2),(0,0),(1,2)\},\ \mathcal{U}^{\text{III-1}}_{1,0}=\{(1,0),(1,1),(2,0)\},\right.\\
&\ \left.\mathcal{U}^{\text{III-1}}_{1,1}=\{(1,1),(1,2),(2,1)\},\ \mathcal{U}^{\text{III-1}}_{1,2}=\{(1,2),(1,0),(2,2)\},\right.\\
&\ \left.\mathcal{U}^{\text{III-1}}_{2,0}=\{(2,0),(2,1),(0,0)\},\ \mathcal{U}^{\text{III-1}}_{2,1}=\{(2,1),(2,2),(0,1)\},\right.\\
&\ \left.\mathcal{U}^{\text{III-1}}_{2,2}=\{(2,2),(2,0),(0,2)\}\right\};
\end{align*}
\begin{align*}
\text{Type III-2:}\ \mathcal{L}_{\text{III-2}}=&\left\{ \mathcal{U}^{\text{III-2}}_{0,0}=\{(0,0),(1,0),(1,1)\},\ \mathcal{U}^{\text{III-2}}_{0,1}=\{(0,1),(1,1),(1,2)\},\right.\\
&\ \left.\mathcal{U}^{\text{III-2}}_{0,2}=\{(0,2),(1,2),(1,0)\},\
\mathcal{U}^{\text{III-2}}_{1,0}=\{(1,0),(2,0),(2,1)\},\right.\\
&\ \left.\mathcal{U}^{\text{III-2}}_{1,1}=\{(1,1),(2,1),(2,2)\},\ \mathcal{U}^{\text{III-2}}_{1,2}=\{(1,2),(2,2),(2,0)\},\right.\\
&\ \left.\mathcal{U}^{\text{III-2}}_{2,0}=\{(2,0),(0,0),(0,1)\},\ \mathcal{U}^{\text{III-2}}_{2,1}=\{(2,1),(0,1),(0,2)\},\right.\\
&\ \left.\mathcal{U}^{\text{III-2}}_{2,2}=\{(2,2),(0,2),(0,0)\}\right\};
\end{align*}
\begin{align*}
\text{Type III-3:}\ \mathcal{L}_{\text{III-3}}=&\left\{ \mathcal{U}^{\text{III-3}}_{0,0}=\{(0,0),(0,1),(1,1)\},\ \mathcal{U}^{\text{III-3}}_{0,1}=\{(0,1),(0,2),(1,2)\},\right.\\
&\ \ \left.\mathcal{U}^{\text{III-3}}_{0,2}=\{(0,2),(0,0),(1,0)\},\ \mathcal{U}^{\text{III-3}}_{1,0}=\{(1,0),(1,1),(2,1)\},\right.\\
&\ \ \left.\mathcal{U}^{\text{III-3}}_{1,1}=\{(1,1),(1,2),(2,2)\},\ \mathcal{U}^{\text{III-3}}_{1,2}=\{(1,2),(1,0),(2,0)\},\right.\\
&\ \ \left.\mathcal{U}^{\text{III-3}}_{2,0}=\{(2,0),(2,1),(0,1)\},\ \mathcal{U}^{\text{III-3}}_{2,1}=\{(2,1),(2,2),(0,2)\},\right.\\
&\ \ \left.\mathcal{U}^{\text{III-3}}_{2,2}=\{(2,2),(2,0),(0,0)\}\right\};
\end{align*}
\begin{align*}
\text{Type III-4:}\ \mathcal{L}_{\text{III-4}}=&\left\{
\mathcal{U}^{\text{III-4}}_{0,0}=\{(0,0),(0,2),(2,0)\},\ \mathcal{U}^{\text{III-4}}_{0,1}=\{(0,1),(0,0),(2,1)\},\right.\\
&\ \ \left.\mathcal{U}^{\text{III-4}}_{0,2}=\{(0,2),(0,1),(2,2)\},\
\mathcal{U}^{\text{III-4}}_{1,0}=\{(1,0),(1,2),(0,0)\},\right.\\
&\ \ \left.\mathcal{U}^{\text{III-4}}_{1,1}=\{(1,1),(1,0),(0,1)\},\ \mathcal{U}^{\text{III-4}}_{1,2}=\{(1,2),(1,1),(0,2)\},\right.\\
&\ \ \left.\mathcal{U}^{\text{III-4}}_{2,0}=\{(2,0),(2,2),(1,0)\},\ \mathcal{U}^{\text{III-4}}_{2,1}=\{(2,1),(2,0),(1,1)\},\right.\\
&\ \ \left.\mathcal{U}^{\text{III-4}}_{2,2}=\{(2,2),(2,1),(1,2)\}\right\}.
\end{align*}

\noindent $\bullet$ From \eqref{eq-type-IV} the user set of Type IV in the red regions of \eqref{eq-type-IV} are
\begin{align*}
\text{Type IV:}\ \mathcal{L}_{\text{IV}}=&\left\{\mathcal{U}^{\text{IV}}_{0,0}=\{(0,0),(0,1),(1,0),(1,1)\},\
\mathcal{U}^{\text{IV}}_{0,1}=\{(0,1),(0,2),(1,1),(1,2)\},\right.\\
&\ \ \left.
\mathcal{U}^{\text{IV}}_{0,2}=\{(0,2),(0,0),(1,2),(1,0)\}, \ \mathcal{U}^{\text{IV}}_{1,0}=\{(1,0),(1,1),(2,0),(2,1)\}, \right.\\
&\ \ \left.\mathcal{U}^{\text{IV}}_{1,1}=\{(1,1),(1,2),(2,1),(2,2)\},\  \mathcal{U}^{\text{IV}}_{1,2}=\{(1,2),(1,0),(2,2),(2,0)\},\right.\\
&\ \ \left.\mathcal{U}^{\text{IV}}_{2,0}=\{(2,0),(2,1),(0,0),(0,1)\},\
\mathcal{U}^{\text{IV}}_{2,1}=\{(2,1),(2,2),(0,1),(0,2)\},\right.\\
&\ \ \left.\mathcal{U}^{\text{IV}}_{2,2}=\{(2,2),(2,0),(0,2),(0,0)\}\right\}.
\end{align*}

By Remark \ref{usertype_r=limit}, if $r=\frac{\sqrt{2}}{2}$, the users consist of $\mathcal{L}_{\text{I}}$ and $\mathcal{L}_{\text{II-1}}$, $\mathcal{L}_{\text{II-2}}$; if $r=1$, the users consist of $\mathcal{L}_{\text{II-1}}$, $\mathcal{L}_{\text{II-2}}$, $\mathcal{L}_{\text{III-1}}$, $\mathcal{L}_{\text{III-2}}$, $\mathcal{L}_{\text{III-3}}$, $\mathcal{L}_{\text{III-4}}$ and $\mathcal{L}_{\text{IV}}$.
\end{example}

\subsection{The First Coded Caching Scheme}
Based on the four types of users introduced in the above subsection, we can obtain our first coded caching Scheme A for a $(K_1, K_2, U,r, M, N)$ 2D caching-aided UDN system by using $(K_1K_2,M,N)$ MN scheme. Specifically we use the placement strategy for users in $(K_1K_2,M,N)$ MN scheme to cache nodes, then use the delivery strategy of $(K_1K_2,M,N)$ MN scheme to each sub-type of users. So Scheme A can be obtained as follows by Algorithm \ref{alg-MN}.

\begin{itemize}
\item \textbf{The placement phase:} By Lines 1-5 of Algorithm \ref{alg-MN}, each file is split into $F={K_1K_2 \choose t}$ packets where $t = \frac{K_1K_2M}{N}\in [0,K_1K_2]$ and each packet is labeled by $\mathcal{T} \in {\mathcal{K} \choose t}$ where $\mathcal{K}=[0:K_1)\times [0:K_2)$. By Lines 6-9 of Algorithm \ref{alg-MN}, each cache node $\text{C}_{k_1,k_2}$ where $(k_1,k_2)\in \mathcal{K}$ caches the packet $W_{n,\mathcal{T}}$ only if $(k_1,k_2)\in \mathcal{T}$, so the cache contents cached by cache node $\text{C}_{k_1,k_2}$ are
    \begin{equation}\label{SC1-Z_Ck}
	\mathcal{Z}_{\text{C}_{k_1,k_2}}=\left\{W_{n,\mathcal{T}}\ \Big|\ (k_1,k_2)\in \mathcal{T}, \mathcal{T}\in {\mathcal{K} \choose t},n \in [0:N)\right\}.
\end{equation}

From \eqref{eq-type-I}, \eqref{eq-type-II}, \eqref{eq-type-III} and \eqref{eq-type-IV}, each user $\mathcal{U}_{k_1,k_2}\in \mathcal{L}$ can retrieve the following packets
\begin{equation}\label{SC1-Z_Uk}
	\mathcal{Z}_{\mathcal{U}_{k_1,k_2}}=\left\{W_{n,\mathcal{T}}\ \Big|\ \mathcal{U}_{k_1,k_2} \bigcap \mathcal{T} \neq \emptyset, \mathcal{T}\in {\mathcal{K}\choose t}, n \in [0,N)\right\}.
\end{equation}
Then we can count that each user of Type I, Type II, Type III and Type IV retrieves ${K_1K_2-1 \choose t-1}$, ${K_1K_2 \choose t}-{K_1K_2-2 \choose t}$, ${K_1K_2 \choose t}-{K_1K_2-3 \choose t}$ and ${K_1K_2 \choose t}-{K_1K_2-4 \choose t}$ different packets of each file respectively. Hence the ratios of the retrieved contents by each user of Type I, Type II, Type III, and Type IV are respectively
\begin{align}
	\label{user_M/N}	
	&\lambda_1=\frac{M_{\mathcal{\text{I}}}}{N}=\frac{{K_1K_2-1 \choose t-1}}{{K_1K_2 \choose t}}=\frac{t}{K_1K_2},\nonumber \\
	&\lambda_2=\frac{M_{\mathcal{\text{II}}}}{N}=\frac{{K_1K_2 \choose t}-{K_1K_2-2 \choose t}}{{K_1K_2 \choose t}}=1-\frac{(K_1K_2-t)(K_1K_2-t-1)}{K_1K_2(K_1K_2-1)},\nonumber \\
	&\lambda_3=\frac{M_{\mathcal{\text{III}}}}{N}=\frac{{K_1K_2 \choose t}-{K_1K_2-3 \choose t}}{{K_1K_2 \choose t}}=1-\frac{(K_1K_2-t)(K_1K_2-t-1)(K_1K_2-t-2)}{K_1K_2(K_1K_2-1)(K_1K_2-2)},\nonumber \\
	&\lambda_4=\frac{M_{\mathcal{\text{IV}}}}{N}=\frac{{K_1K_2 \choose t}-{K_1K_2-4 \choose t}}{{K_1K_2 \choose t}}\nonumber \\
	&\ \ \  =1-\frac{(K_1K_2-t)(K_1K_2-t-1)(K_1K_2-t-2)(K_1K_2-t-3)}{K_1K_2(K_1K_2-1)(K_1K_2-2)(K_1K_2-3)}.
\end{align}
\item \textbf{The delivery phase:} Each user randomly requests a file from the server. Then for each sub-type of users, we use the delivery strategy in Lines 10-12 of Algorithm \ref{alg-MN} to send all the ${K_1K_2 \choose t+1}$ coded signals to the users.
\end{itemize}

From \eqref{eq:MN}, the transmission load is $\frac{K_1K_2-t}{t+1}$ for each sub-type. Then, we use the above transmission strategies for each sub-type of user according to \eqref{allusers}, \eqref{allusers_min} and \eqref{allusers_max}, the following result can be obtained.

\begin{theorem}[Scheme A]\rm
\label{th-baseline}	
For any positive integers $K_1$, $K_2$, $M$ and $N$ where $K_1,K_2\geq 3$ and $K_1K_2M/N\in [0:K_1K_2]$, there exists a coded caching scheme for $(K_1, K_2, U,r, M, N)$ 2D caching-aided UDN system with transmission load
\begin{equation*}
	R_{\text{A}}=
	\left\{
	\begin{array}{lr}
		3 \cdot \frac{(K_1K_2-t)}{t+1}, \qquad \ \ \ \text{if }\ r=\frac{\sqrt{2}}{2},\\
		8\cdot \frac{(K_1K_2-t)}{t+1}, \qquad \ \ \ \text{if }\ \frac{\sqrt{2}}{2}< r < 1,\\
		7\cdot \frac{(K_1K_2-t)}{t+1}, \qquad \ \ \ \text{if }\ r=1.
	\end{array}
	\right.
\end{equation*}
Moreover, if $K_1K_2\rightarrow \infty$, we have
\begin{equation}
	\label{eq-load-baseline_all-approx}	
	R_{\text{A}}\approx
	\left\{
	\begin{array}{lr}
		3\cdot \frac{N}{M}\left(1-\frac{M}{N}\right), \qquad \text{if }\ r=\frac{\sqrt{2}}{2},\\
		8\cdot \frac{N}{M}\left(1-\frac{M}{N}\right), \qquad \text{if }\ \frac{\sqrt{2}}{2}< r < 1,\\
		7\cdot \frac{N}{M}\left(1-\frac{M}{N}\right), \qquad \text{if }\ r=1.
	\end{array}
	\right.
\end{equation}
\hfill $\square$
\end{theorem}
\begin{remark}[The optimal delivery strategy for the users of Type I]\rm
In Type I, each user accesses exactly the cache node. This implies that each user takes a unique cache node as its own memory. In addition, the delivery strategy is the same as that of the MN scheme too. So Scheme A for the users in Type I is equivalent to the $(K_1K_2,M,M)$ MN scheme. It is well known that the MN scheme has the minimum transmission load under uncoded placement when $K_1K_2\leq N$. Hence Scheme A for the users in Type I also has the minimum transmission load.
\end{remark}

\subsection{Performance Analyses}
In fact, our scheme can be regarded as a special scheme with heterogenous memory sizes for the shared-link caching system. In a $(K,\mathcal{M},N)$ shared-link coded caching system, let user $k\in [0:K)$ has a memory with size $M_k$. Without loss of generality, we assume that $M_0\leq M_1 \leq \ldots \leq M_{K-1}$. In the following we will discuss the performance of our scheme compared with the order-optimal scheme proposed in \cite{WSWXH} which is listed as follows.

\begin{lemma}[Order optimality of scheme in \cite{WSWXH}]\rm
\label{lemma:perf-wswxh}
There exists an order optimal scheme for a $(K,\mathcal{M},N)$ shared-link coded caching system with transmission load
\begin{align*}
R_{\text{D}}=\sum_{i=0}^{K-1}\left[\prod_{j=0}^{i}\left(1-\frac{M_j}{N}\right)\right]
\end{align*}
which is in a multiplicative gap of at most $6$, i.e.,
\begin{align}
	\label{compare:R_D,R*}
\frac{R_{\text{D}}}{R^*(K,\mathcal{M},N)}\leq6.
\end{align}
\end{lemma}

Now let consider our Scheme A in $(K_1, K_2, U,r, M, N)$ 2D caching-aided UDN system where $\frac{\sqrt{2}}{2} < r < 1$. From \eqref{user_M/N}, there are $8K_1K_2$ users.
We can regard our Scheme A as a $(8K_1K_2,\mathcal{M}, N)$ shared link coded caching scheme where
\begin{align*}
\mathcal{M}&=\{\underbrace{M_{\mathcal{\text{I}}},M_{\mathcal{\text{I}}},\ldots,
M_{\mathcal{\text{I}}}}_{K_1K_2 \ \text{users of Type I}},\
\underbrace{M_{\mathcal{\text{II}}},M_{\mathcal{\text{II}}},\ldots,
M_{\mathcal{\text{II}}}}_{2K_1K_2 \ \text{users of Type II}},\
\underbrace{M_{\mathcal{\text{III}}},M_{\mathcal{\text{III}}},\ldots,
M_{\mathcal{\text{III}}}}_{4K_1K_2 \ \text{users of Type III}},\
\underbrace{M_{\mathcal{\text{IV}}},M_{\mathcal{\text{IV}}},\ldots,
M_{\mathcal{\text{IV}}}}_{K_1K_2 \ \text{users of Type IV}}
\}\\
&=\{M_0,M_1,\ldots,M_{K_1K_2-1}\}.
\end{align*}
Here $M_{\mathcal{\text{I}}}=\lambda_1 N$, $M_{\mathcal{\text{II}}}=\lambda_2 N$, $M_{\mathcal{\text{III}}}=\lambda_3 N$ and $M_{\mathcal{\text{IV}}}=\lambda_4 N$ where $\lambda_1$, $\lambda_2$, $\lambda_3$, $\lambda_4$ are defined in \eqref{user_M/N}. By Lemma \ref{lemma:perf-wswxh}, we have a scheme with the transmission load
\begin{align}
\label{R_D-8K}
R_{\text{D}}&=\sum_{i=0}^{8K_1K_2-1}\left[\prod_{j=0}^{i}\left(1-\frac{M_j}{N}\right)\right]\nonumber\\
&=\left(\frac{1}{\lambda_1}-1\right)\left[1-(1-\lambda_1)^{K_1K_2}\right]+(1-\lambda_1)^{K_1K_2}
\left(\frac{1}{\lambda_2}-1\right)\left[1-(1-\lambda_2)^{2K_1K_2}\right]\nonumber\\
&\ \ \ +(1-\lambda_1)^{K_1K_2}(1-\lambda_2)^{2K_1K_2}\left(\frac{1}{\lambda_3}-1\right)
\left[1-(1-\lambda_3)^{4K_1K_2}\right]\nonumber\\
&\ \ \ +(1-\lambda_1)^{K_1K_2}(1-\lambda_2)^{2K_1K_2}(1-\lambda_3)^{4K_1K_2}
\left(\frac{1}{\lambda_4}-1\right)\left[1-(1-\lambda_4)^{K_1K_2}\right]
\nonumber\\
&\approx \left(\frac{1}{\lambda_1}-1\right), \ \ \ \ \ \ \ \ \ (K_1K_2\rightarrow +\infty).
\end{align}
Then from \eqref{eq-load-baseline_all-approx} and \eqref{R_D-8K} we have
\begin{align*}
\frac{R_{\text{A}}}{R_{\text{D}}}\approx 8.
\end{align*}
In addition, from \eqref{compare:R_D,R*} we have
\begin{align}
\label{order-optimal-1}
\frac{R_{\text{A}}}{R^*(K,\mathcal{M},N)}\leq \frac{6R_{\text{A}}}{R_{\text{D}}}\leq 48.
\end{align}

Similar to the above discussion, the following results can be obtained directly.
\begin{eqnarray}
\label{order-optimal-2}
\begin{split}
\left\{\begin{array}{cc}
\frac{R_{\text{A}}}{R^*(K,\mathcal{M}',N)}\leq 18,&\ \ \text{if}\ r=\frac{\sqrt{2}}{2},\\
\frac{R_{\text{A}}}{R^*(K,\mathcal{M}'',N)}\leq 42,& \text{if}\ r=1,
\end{array}\right.\ \ \ \ \ \ \ \  (K_1K_2\rightarrow +\infty).
\end{split}
\end{eqnarray}  where
\begin{align*}
&\mathcal{M}'=\{\underbrace{M_{\mathcal{\text{I}}},M_{\mathcal{\text{I}}},\ldots,
M_{\mathcal{\text{I}}}}_{K_1K_2 \ \text{users of Type I}},\
\underbrace{M_{\mathcal{\text{II}}},M_{\mathcal{\text{II}}},\ldots,
M_{\mathcal{\text{II}}}}_{2K_1K_2 \ \text{users of Type II}}
\},\\
&\mathcal{M}''=\{
\underbrace{M_{\mathcal{\text{II}}},M_{\mathcal{\text{II}}},\ldots,
M_{\mathcal{\text{II}}}}_{2K_1K_2 \ \text{users of Type II}},\
\underbrace{M_{\mathcal{\text{III}}},M_{\mathcal{\text{III}}},\ldots,
M_{\mathcal{\text{III}}}}_{4K_1K_2 \ \text{users of Type III}},\
\underbrace{M_{\mathcal{\text{IV}}},M_{\mathcal{\text{IV}}},\ldots,
M_{\mathcal{\text{IV}}}}_{K_1K_2 \ \text{users of Type IV}}
\}.
\end{align*}
From \eqref{order-optimal-1} and \eqref{order-optimal-2}, the proposed Scheme A is order optimal.

\section{Improve scheme B by reducing the redundancy in Scheme A}
\label{sec-Improved}
In this section, we investigate the redundant transmissions in the delivery phase of Scheme A for the users of Type II, Type III and Type IV respectively, and then obtain the following Scheme B which further reduces the amount of transmission overhead of Scheme A.

\subsection{Research Motivation}
Let us consider the delivery strategy of Scheme A in Section \ref{sec-baseline}. Then the following investigations can be obtained. In Type II, each user can access two different cache nodes. In this case, we claim that there are
$h^{\text{II}}_{1}={K_1K_2-1 \choose t}-{K_1K_2-2\choose t}$ coded signals, and
$h^{\text{II}}_{2}={K_1K_2-2 \choose t}-{K_1K_2-3\choose t}$ modified coded signals
in Line 11 of Algorithm \ref{alg-MN} which can be retrieved by each user from its accessing cache nodes. So there are
\begin{align}
\label{eq-redundancy-II}
h^{\text{II}}=h^{\text{II}}_{1}+h^{\text{II}}_{2}={K_1K_2-1 \choose t}-{K_1K_2-3\choose t}
\end{align} coded signals (or modified coded) which can are retrieved by user $\mathcal{U}^{\text{II}}_{k_1,k_2}$.

Let's explain the above statement. First, we consider the users in Type II-1. For any integer pair $(k_1,k_2)\in\mathcal{K}$, user $\mathcal{U}^{\text{II-1}}_{k_1,k_2}=\{(k_1,k_2),(k_1,<k_2+1>_{K_2})\}$ can access cache nodes $\text{C}_{k_1,k_2}$ and $\text{C}_{k_1,<k_2+1>_{K_2}}$, and user $\mathcal{U}^{\text{II-1}}_{k_1,<k_2+1>_{K_2}}=\{(k_1,<k_2+1>_{K_2}),(k_1,<k_2+2>_{K_2})\}$ can access cache nodes $\text{C}_{k_1,<k_2+1>_{K_2}}$ and $\text{C}_{k_1,<k_2+2>_{K_2}}$. Let us consider the delivery strategy for users $\mathcal{U}^{\text{II-1}}_{k_1,k_2}$ and $\mathcal{U}^{\text{II-1}}_{k_1,<k_2+1>_{K_2}}$ of Scheme A by Lines 10-12 of Algorithm \ref{alg-MN}.
\begin{itemize}
\item First we consider the value of $h^{\text{II}}_{1}$ by counting the number of $t$-subset $\mathcal{T}_1\in {\mathcal{K} \choose t}$ satisfying the conditions
\begin{align}
\label{cond:II-1_T1}
(k_1,k_2) \notin \mathcal{T}_1,\ \text{and }(k_1,<k_2+1>_{K_2}) \in \mathcal{T}_1.
\end{align}
Let $\mathcal{S}_1=\{(k_1,k_2)\}\bigcup \mathcal{T}_1$. From \eqref{eq-type-II-1} and \eqref{SC1-Z_Uk}, we have $(\mathcal{S}_1\setminus\{(k_1',k_2')\})\bigcap \mathcal{U}^{\text{II-1}}_{k_1,k_2}\neq \emptyset$ where $(k_1',k_2')\in \mathcal{S}_1$, then the transmitted coded signal
$X_{\mathcal{S}_1}=\bigoplus_{(k_1',k_2')\in \mathcal{S}_1}W_{d_{\mathcal{U}_{k_1',k_2'}},\mathcal{S}_1\setminus\{(k_1',k_2')\}}$
can be retrieved by user $\mathcal{U}^{\text{II-1}}_{k_1,k_2}$, i.e., the packets decoded from $X_{\mathcal{S}_1}$ can be retrieved by user $\mathcal{U}^{\text{II-1}}_{k_1,k_2}$ through cache nodes $\text{C}_{k_1,k_2}$ and $\text{C}_{k_1,<k_2+1>_{K_2}}$. We can check that there are $h^{\text{II}}_{1}={K_1K_2-1 \choose t}-{K_1K_2-2 \choose t}$ such subsets $\mathcal{T}_1$. So there are $h^{\text{II}}_{1}$ such coded signals in Line 11 of Algorithm \ref{alg-MN} which can be retrieved by user $\mathcal{U}^{\text{II-1}}_{k_1,k_2}$ from its accessing cache nodes.
\item Then we consider the value of $h^{\text{II}}_{2}$ by counting the number of $t$-subset $\mathcal{T}_2\in {\mathcal{K} \choose t}$ satisfying the conditions
\begin{align}
\label{cond:II-1_T2}
(k_1,k_2) \notin \mathcal{T}_2,\ (k_1,<k_2+1>_{K_2}) \notin \mathcal{T}_2,\ \text{and }(k_1,<k_2+2>_{K_2}) \in \mathcal{T}_2.
\end{align}
Let $\mathcal{S}_2=\{(k_1,<k_2+1>_{K_2})\}\bigcup \mathcal{T}_2$. From \eqref{eq-type-II-1} and \eqref{SC1-Z_Uk}, we have $(\mathcal{S}_2\setminus\{(k_1',k_2')\})\bigcap$ $\mathcal{U}^{\text{II-1}}_{k_1,<k_2+1>_{K_2}} \neq \emptyset$ where $(k_1',k_2')\in \mathcal{S}_2$, then the transmitted coded signal
	$X_{\mathcal{S}_2}=\bigoplus_{(k_1',k_2')\in \mathcal{S}_2}$ $W_{d^{\text{II-1}}_{\mathcal{U}_{k_1',k_2'}},\mathcal{S}_2\setminus\{(k_1',k_2')\}}$
can be retrieved by $\mathcal{U}^{\text{II-1}}_{k_1,<k_2+1>_{K_2}}$, i.e., the packets contained in $X_{\mathcal{S}_2}$ can be retrieved by user $\mathcal{U}^{\text{II-1}}_{k_1,<k_2+1>_{K_2}}$ through cache nodes $\text{C}_{k_1,<k_2+1>_{K_2}}$ and $\text{C}_{k_1,<k_2+2>_{K_2}}$. Then the server only needs to send the following modified coded signal
$X'_{\mathcal{S}_2}=X_{\mathcal{S}_2}-W_{d_{\mathcal{U}^{\text{II-1}}_{k_1,<k_2+1>_{K_2}}},\mathcal{T}_2}=\bigoplus_{(k'_1,k'_2)\in \mathcal{S}_2\setminus\{(k_1,<k_2+1>_{K_2})\}}W_{d_{\mathcal{U}^{\text{II-1}}_{k'_1,k'_2}},\mathcal{S}_2\setminus\{(k'_1,k'_2)\}}$.
From \eqref{SC1-Z_Uk}, we can check that user $\mathcal{U}^{\text{II-1}}_{k_1,k_2}$ can also retrieve all the packets in $X'_{\mathcal{S}_2}$ since it can retrieve all the packets labeled by $\mathcal{S}_2\setminus\{(k'_1,k'_2)\}$ where $(k'_1,k'_2)\neq (k_1,<k_2+1>_{K_2})$. We can check that there are $h^{\text{II}}_{2}={K_1K_2-2 \choose t}-{K_1K_2-3\choose t}$ such subsets $\mathcal{T}_2$. So there are $h^{\text{II}}_{2}$ such modified coded signals in Line 11 of Algorithm \ref{alg-MN} which can be retrieved by user $\mathcal{U}^{\text{II-1}}_{k_1,k_2}$ from its accessing cache nodes.
\end{itemize}

Finally, we claim that $\mathcal{S}_1\neq\mathcal{S}_2$ holds since $(k_1,k_2) \in \mathcal{S}_1$ and $(k_1,k_2) \notin \mathcal{S}_2$. So there are exactly $h^{\text{II}}=h^{\text{II}}_{1}+h^{\text{II}}_{2}$ coded (or modified coded) signals which can be retrieved by user $\mathcal{U}^{\text{II-1}}_{k_1,k_2}$.

Similarly, for each Type II-2 user $\mathcal{U}^{\text{II-2}}_{k_1,k_2}=\{(k_1,k_2),(<k_1+1>_{K_1},k_2)\}$ and user $\mathcal{U}^{\text{II-2}}_{<k_1+1>_{K_1},k_2}$ $=\{(<k_1+1>_{K_1},k_2),(<k_1+2>_{K_1},k_2)\}$, by studying the cache contents in the cache nodes $\text{C}_{k_1,k_2}$ and $\text{C}_{<k_1+1>_{K_1},k_2}$ accessed by user $\mathcal{U}^{\text{II-2}}_{k_1,k_2}$ and the cache contents in the cache nodes $\text{C}_{<k_1+1>_{K_1},k_2}$ and $\text{C}_{<k_1+2>_{K_1},k_2}$ accessed by user $\mathcal{U}^{\text{II-2}}_{<k_1+1>_{K_1},k_2}$, there are also
$h^{\text{II}}_{1}={K_1K_2-1 \choose t}-{K_1K_2-2\choose t}$ coded signals and
$h^{\text{II}}_{2}={K_1K_2-2 \choose t}-{K_1K_2-3\choose t}$ modified coded signals which can be retrieved by user $\mathcal{U}^{\text{II-2}}_{k_1,k_2}$ from its accessing cache nodes.

Similar to the discussion of Type II, according to the packets retrieved by the users in Type III and Type IV and the modified transmitted coded signals of Scheme A, the following statement can be obtained.
There are
$h^{\text{III}}_1={K_1K_2-1 \choose t}-{K_1K_2-3\choose t}$ coded signals and
$h^{\text{III}}_2={K_1K_2-2 \choose t}-{K_1K_2-4\choose t}$ modified coded signals,
$h^{\text{III}}_3={K_1K_2-3 \choose t}-{K_1K_2-5\choose t}$ modified coded signals
in Line 11 of Algorithm \ref{alg-MN} which can be retrieved by each user of Type III. And then there are
\begin{align}
\label{eq-redundancy-III}
h^{\text{III}}=h^{\text{III}}_{1}+h^{\text{III}}_{2}+h^{\text{III}}_{3}=
\sum\limits_{i=1}^{2}{K_1K_2-i \choose t}-\sum\limits_{i'=4}^{5}{K_1K_2-i' \choose t}
\end{align}
coded signals (or modified coded) which are retrieved by each user of Type III.
There are
$h^{\text{IV}}_1={K_1K_2-1 \choose t}-{K_1K_2-4\choose t}$ coded signals and
$h^{\text{IV}}_2={K_1K_2-2 \choose t}-{K_1K_2-5\choose t}$ modified coded signals,
$h^{\text{IV}}_3={K_1K_2-3 \choose t}-{K_1K_2-6\choose t}$ modified coded signals,
$h^{\text{IV}}_4={K_1K_2-4 \choose t}-{K_1K_2-7\choose t}$ modified coded signals
in Line 11 of Algorithm \ref{alg-MN} which can be retrieved by each user of Type IV. And then there are
\begin{align}
h^{\text{IV}}&=h^{\text{IV}}_{1}+h^{\text{IV}}_{2}+h^{\text{IV}}_{3}+h^{\text{IV}}_{4}=
\sum\limits_{i=1}^{3}{K_1K_2-i \choose t}
-\sum\limits_{i'=5}^{7}{K_1K_2-i'\choose t}\label{eq-redundancy-IV}
\end{align}
coded signals (or modified coded) which are retrieved by each Type III user.

The detailed computations of $h^{\text{III}}_1$, $h^{\text{III}}_2$, $h^{\text{III}}_3$ and $h^{\text{IV}}_1$, $h^{\text{IV}}_2$, $h^{\text{IV}}_3$, $h^{\text{IV}}_4$ for Type III user and Type IV can be found in Appendix \ref{appendix-Type III_h} and \ref{appendix-Type IV_h} respectively.

In fact, we can use a maximum distance separable (MDS) code to reduce the redundant coded (and modified) signals whose sizes are listed in \eqref{eq-redundancy-II}, \eqref{eq-redundancy-III} and \eqref{eq-redundancy-IV} respectively for each user of Type II, Type III and Type IV. For instance, by a $\left[{K_1K_2\choose t+1},{K_1K_2\choose t+1}-h^{\text{II}}\right]_ q$ MDS code with some prime power $q$, the server only needs to transmit ${K_1K_2\choose t+1}-h^{\text{II}}$ coded signals for the the users in Type II-1. So we encode the ${K_1K_2\choose t+1}$ transmitted coded signals for the the users in Type II-2, Type III-1, Type III-2, Type III-3, Type III-4, Type IV respectively by $\left[{K_1K_2\choose t+1},{K_1K_2\choose t+1}-h^{\text{II}}\right]_q$ MDS code, $\left[{K_1K_2\choose t+1},{K_1K_2\choose t+1}-h^{\text{III}}\right]_ q$ MDS code and $\left[{K_1K_2\choose t+1},{K_1K_2\choose t+1}-h^{\text{IV}}\right]_q$ MDS code with some prime power $q$. Then when $\frac{\sqrt{2}}{2} \leq r \leq 1$, the total reduction of transmitted coded signals in Scheme A can be obtained respectively.
\begin{equation*}
	H=
	\left\{
	\begin{array}{lr}
		2h^{\text{II}}, \qquad \qquad \qquad \qquad \qquad \text{if } r=\frac{\sqrt{2}}{2},\\
		2h^{\text{II}}+4h^{\text{III}}+h^{\text{IV}},\qquad \qquad  \ \text{if } \frac{\sqrt{2}}{2}< r \leq 1.
	\end{array}
	\right.
\end{equation*}

From \eqref{eq-load-baseline_all-approx} and the above equation, we can calculate the transmission load of Scheme B, as follows.
\begin{itemize}
\item When $r=\frac{\sqrt{2}}{2}$, the transmission load is
\begin{align*}
R_{\text{B}}&= \frac{3S-H}{F}= \frac{3{K_1K_2 \choose t+1}-2{K_1K_2-1 \choose t}+2{K_1K_2-3\choose t}}{{K_1K_2 \choose t}}\\
&=\frac{3{K_1K_2-1 \choose t+1}+{K_1K_2-1 \choose t}+2{K_1K_2-3\choose t}}{{K_1K_2 \choose t}}\\
&=\frac{3(K_1K_2-t)(K_1K_2-t-1)}{K_1K_2(t+1)}+\frac{K_1K_2-t}{K_1K_2}
+\frac{2(K_1K_2-t)(K_1K_2-t-1)}{K_1K_2(K_1K_2-1)} \\
&\ \ \ \ \cdot \frac{K_1K_2-t-2}{K_1K_2-2} \\
&=3\left(1-\frac{M}{N}\right)\left(1-\frac{M}{N}-\frac{1}{K_1K_2}\right) \frac{1}{\frac{M}{N}+\frac{1}{K_1K_2}}+\left(1-\frac{M}{N}\right)+2(1-\frac{M}{N})\\
&\ \ \ \ \cdot \frac{(1-\frac{M}{N}-\frac{1}{K_1K_2})(1-\frac{M}{N}-\frac{2}{K_1K_2})}{(1-\frac{1}{K_1K_2})(1-\frac{2}{K_1K_2})}\\
&\approx3\frac{N}{M}\left(1-\frac{M}{N}\right)^2+\left(1-\frac{M}{N}\right)+2\left(1-\frac{M}{N}\right)^3,\ \ \ \ \ \ \ \ \ (K_1K_2\rightarrow \infty).
\end{align*}

\item When $\frac{\sqrt{2}}{2}< r < 1$, the transmission load is
\begin{align*}
R_{\text{B}}&= \frac{8S-H}{F}\\
&=\frac{8{K_1K_2 \choose t+1}-7{K_1K_2-1 \choose t}-5{K_1K_2-2\choose t}+{K_1K_2-3 \choose t}+4{K_1K_2-4 \choose t}+5{K_1K_2-5 \choose t}}{{K_1K_2 \choose t}}\\
&\ \ \ \ + \frac{{K_1K_2-6 \choose t}+{K_1K_2-7 \choose t}}{{K_1K_2 \choose t}}\\
&=\frac{8{K_1K_2-1 \choose t+1}+{K_1K_2-1 \choose t}-5{K_1K_2-2\choose t}+{K_1K_2-3 \choose t}+4{K_1K_2-4 \choose t}+5{K_1K_2-5 \choose t}}{{K_1K_2 \choose t}}\\
&\ \ \ \ + \frac{{K_1K_2-6 \choose t}+{K_1K_2-7 \choose t}}{{K_1K_2 \choose t}}\\
&=\frac{8(K_1K_2-t)(K_1K_2-t-1))}{K_1K_2(t+1)}+\frac{K_1K_2-t}{K_1K_2}-\frac{5(K_1K_2-t)(K_1K_2-t-1)}{(K_1K_2)(K_1K_2-1)}\\
&\ \ \ \ +\frac{(K_1K_2-t)(K_1K_2-t-1)(K_1K_2-t-2)}{(K_1K_2)(K_1K_2-1)(K_1K_2-2)}\\
&\ \ \ \ +\frac{4(K_1K_2-t)(K_1K_2-t-1)(K_1K_2-t-2)(K_1K_2-t-3)}{(K_1K_2)(K_1K_2-1)(K_1K_2-2)(K_1K_2-3)}\\
&\ \ \ \ +\frac{5(K_1K_2-t)(K_1K_2-t-1)(K_1K_2-t-2)(K_1K_2-t-3)(K_1K_2-t-4)}{(K_1K_2)(K_1K_2-1)(K_1K_2-2)(K_1K_2-3)(K_1K_2-4)}\\
&\ \ \ \
+\frac{(K_1K_2-t)(K_1K_2-t-1)(K_1K_2-t-2)(K_1K_2-t-3)(K_1K_2-t-4)}{(K_1K_2)(K_1K_2-1)(K_1K_2-2)(K_1K_2-3)(K_1K_2-4)}\\
&\ \ \ \  \cdot \frac{K_1K_2-t-5}{K_1K_2-5} \\
&\ \ \ \
+\frac{(K_1K_2-t)(K_1K_2-t-1)(K_1K_2-t-2)(K_1K_2-t-3)(K_1K_2-t-4)}{(K_1K_2)(K_1K_2-1)(K_1K_2-2)(K_1K_2-3)(K_1K_2-4)}\\
&\ \ \ \  \cdot \frac{(K_1K_2-t-5)(K_1K_2-t-6)}{(K_1K_2-5)(K_1K_2-6)} \\
&=8\left(1-\frac{M}{N}\right)\left(1-\frac{M}{N}-\frac{1}{K_1K_2}\right) \frac{1}{\frac{M}{N}+\frac{1}{K_1K_2}}+\left(1-\frac{M}{N}\right)\\
&\ \ \ \ -\frac{5(1-\frac{M}{N})(1-\frac{M}{N}-\frac{1}{K_1K_2})}{1-\frac{1}{K_1K_2}}\\
&\ \ \ \ +\frac{(1-\frac{M}{N})(1-\frac{M}{N}-\frac{1}{K_1K_2})(1-\frac{M}{N}-\frac{2}{K_1K_2})}{(1-\frac{1}{K_1K_2})(1-\frac{2}{K_1K_2})}\\
&\ \ \ \ +\frac{4(1-\frac{M}{N})(1-\frac{M}{N}-\frac{1}{K_1K_2})(1-\frac{M}{N}-\frac{2}{K_1K_2})(1-\frac{M}{N}-\frac{3}{K_1K_2})}{(1-\frac{1}{K_1K_2})(1-\frac{2}{K_1K_2})(1-\frac{3}{K_1K_2})}\\
&\ \ \ \ +\frac{5(1-\frac{M}{N})(1-\frac{M}{N}-\frac{1}{K_1K_2})(1-\frac{M}{N}-\frac{2}{K_1K_2})(1-\frac{M}{N}-\frac{3}{K_1K_2})(1-\frac{M}{N}-\frac{4}{K_1K_2})}{(1-\frac{1}{K_1K_2})(1-\frac{2}{K_1K_2})(1-\frac{3}{K_1K_2})(1-\frac{4}{K_1K_2})}\\
&\ \ \ \ +\frac{(1-\frac{M}{N})(1-\frac{M}{N}-\frac{1}{K_1K_2})(1-\frac{M}{N}-\frac{2}{K_1K_2})(1-\frac{M}{N}-\frac{3}{K_1K_2})(1-\frac{M}{N}-\frac{4}{K_1K_2})}{(1-\frac{1}{K_1K_2})(1-\frac{2}{K_1K_2})(1-\frac{3}{K_1K_2})(1-\frac{4}{K_1K_2})}\\
&\ \ \ \ \cdot \frac{1-\frac{M}{N}-\frac{5}{K_1K_2}}{1-\frac{5}{K_1K_2}}\\
&\ \ \ \ +\frac{(1-\frac{M}{N})(1-\frac{M}{N}-\frac{1}{K_1K_2})(1-\frac{M}{N}-\frac{2}{K_1K_2})(1-\frac{M}{N}-\frac{3}{K_1K_2})(1-\frac{M}{N}-\frac{4}{K_1K_2})}{(1-\frac{1}{K_1K_2})(1-\frac{2}{K_1K_2})(1-\frac{3}{K_1K_2})(1-\frac{4}{K_1K_2})}\\
&\ \ \ \ \cdot \frac{(1-\frac{M}{N}-\frac{5}{K_1K_2})(1-\frac{M}{N}-\frac{6}{K_1K_2})}{(1-\frac{5}{K_1K_2})(1-\frac{6}{K_1K_2})} \\
&\approx \left(8\frac{N}{M}-6\right)\left(1-\frac{M}{N}\right)^2+3\left(1-\frac{M}{N}\right)^4+4\left(1-\frac{M}{N}\right)^5+\sum\limits_{i=1}^{7}\left(1-\frac{M}{N}\right)^i,\\
&\ \ \ \ (K_1K_2\rightarrow \infty).
\end{align*}
\item When $r = 1$, the transmission load is
\begin{align*}
R_{\text{B}}&= \frac{7S-H}{F}\\
&=\frac{7{K_1K_2 \choose t+1}-7{K_1K_2-1 \choose t}-5{K_1K_2-2\choose t}+{K_1K_2-3 \choose t}+4{K_1K_2-4 \choose t}+5{K_1K_2-5 \choose t}}{{K_1K_2 \choose t}}\\
&\ \ \ \ +\frac{{K_1K_2-6 \choose t}+{K_1K_2-7 \choose t}}{{K_1K_2 \choose t}}\\
&=\frac{7{K_1K_2-1 \choose t+1}-5{K_1K_2-2\choose t}+{K_1K_2-3 \choose t}+4{K_1K_2-4 \choose t}+5{K_1K_2-5 \choose t}+{K_1K_2-6 \choose t}}{{K_1K_2 \choose t}}\\
&\ \ \ \ +\frac{{K_1K_2-7 \choose t}}{{K_1K_2 \choose t}}\\
&=\frac{7(K_1K_2-t)(K_1K_2-t-1))}{K_1K_2(t+1)}-\frac{5(K_1K_2-t)(K_1K_2-t-1)}{(K_1K_2)(K_1K_2-1)}\\
&\ \ \ \ +\frac{(K_1K_2-t)(K_1K_2-t-1)(K_1K_2-t-2)}{(K_1K_2)(K_1K_2-1)(K_1K_2-2)}\\
&\ \ \ \ +\frac{4(K_1K_2-t)(K_1K_2-t-1)(K_1K_2-t-2)(K_1K_2-t-3)}{(K_1K_2)(K_1K_2-1)(K_1K_2-2)(K_1K_2-3)}\\
&\ \ \ \ +\frac{5(K_1K_2-t)(K_1K_2-t-1)(K_1K_2-t-2)(K_1K_2-t-3)(K_1K_2-t-4)}{(K_1K_2)(K_1K_2-1)(K_1K_2-2)(K_1K_2-3)(K_1K_2-4)}\\
&\ \ \ \
+\frac{(K_1K_2-t)(K_1K_2-t-1)(K_1K_2-t-2)(K_1K_2-t-3)(K_1K_2-t-4)}{(K_1K_2)(K_1K_2-1)(K_1K_2-2)(K_1K_2-3)(K_1K_2-4)}\\
&\ \ \ \ \cdot \frac{K_1K_2-t-5}{K_1K_2-5}  \\
&\ \ \ \
+\frac{(K_1K_2-t)(K_1K_2-t-1)(K_1K_2-t-2)(K_1K_2-t-3)(K_1K_2-t-4)}{(K_1K_2)(K_1K_2-1)(K_1K_2-2)(K_1K_2-3)(K_1K_2-4)}\\
&\ \ \ \ \cdot \frac{(K_1K_2-t-5)(K_1K_2-t-6)}{(K_1K_2-5)(K_1K_2-6)} \\
&=7\left(1-\frac{M}{N}\right)\left(1-\frac{M}{N}-\frac{1}{K_1K_2}\right) \frac{1}{\frac{M}{N}+\frac{1}{K_1K_2}}-\frac{5(1-\frac{M}{N})(1-\frac{M}{N}-\frac{1}{K_1K_2})}{1-\frac{1}{K_1K_2}}\\
&\ \ \ \ +\frac{(1-\frac{M}{N})(1-\frac{M}{N}-\frac{1}{K_1K_2})(1-\frac{M}{N}-\frac{2}{K_1K_2})}{(1-\frac{1}{K_1K_2})(1-\frac{2}{K_1K_2})}\\
&\ \ \ \ +\frac{4(1-\frac{M}{N})(1-\frac{M}{N}-\frac{1}{K_1K_2})(1-\frac{M}{N}-\frac{2}{K_1K_2})(1-\frac{M}{N}-\frac{3}{K_1K_2})}{(1-\frac{1}{K_1K_2})(1-\frac{2}{K_1K_2})(1-\frac{3}{K_1K_2})}\\
&\ \ \ \ +\frac{5(1-\frac{M}{N})(1-\frac{M}{N}-\frac{1}{K_1K_2})(1-\frac{M}{N}-\frac{2}{K_1K_2})(1-\frac{M}{N}-\frac{3}{K_1K_2})(1-\frac{M}{N}-\frac{4}{K_1K_2})}{(1-\frac{1}{K_1K_2})(1-\frac{2}{K_1K_2})(1-\frac{3}{K_1K_2})(1-\frac{4}{K_1K_2})}\\
&\ \ \ \ +\frac{(1-\frac{M}{N})(1-\frac{M}{N}-\frac{1}{K_1K_2})(1-\frac{M}{N}-\frac{2}{K_1K_2})(1-\frac{M}{N}-\frac{3}{K_1K_2})(1-\frac{M}{N}-\frac{4}{K_1K_2})}{(1-\frac{1}{K_1K_2})(1-\frac{2}{K_1K_2})(1-\frac{3}{K_1K_2})(1-\frac{4}{K_1K_2})}\\
&\ \ \ \ \cdot \frac{1-\frac{M}{N}-\frac{5}{K_1K_2}}{1-\frac{5}{K_1K_2}} \\
&\ \ \ \ +\frac{(1-\frac{M}{N})(1-\frac{M}{N}-\frac{1}{K_1K_2})(1-\frac{M}{N}-\frac{2}{K_1K_2})(1-\frac{M}{N}-\frac{3}{K_1K_2})(1-\frac{M}{N}-\frac{4}{K_1K_2})}{(1-\frac{1}{K_1K_2})(1-\frac{2}{K_1K_2})(1-\frac{3}{K_1K_2})(1-\frac{4}{K_1K_2})}\\
&\ \ \ \ \cdot \frac{(1-\frac{M}{N}-\frac{5}{K_1K_2})(1-\frac{M}{N}-\frac{6}{K_1K_2})}{(1-\frac{5}{K_1K_2})(1-\frac{6}{K_1K_2})} \\
&\approx \left(7\frac{N}{M}-6\right)\left(1-\frac{M}{N}\right)^2+3\left(1-\frac{M}{N}\right)^4+4\left(1-\frac{M}{N}\right)^5+\sum\limits_{i=2}^{7}\left(1-\frac{M}{N}\right)^i,\\
&\quad \ (K_1K_2\rightarrow \infty).
\end{align*}
\end{itemize}

With the scheme described above, we achieve the transmission load in the following Theorem.

\begin{theorem}[Scheme B] \rm \label{thm-scheme1}
For any positive integers $K_1$, $K_2$, $M$, $N$ and $t=K_1K_2M/N\in [0,K_1K_2]$ where $K_1,K_2\geq 3$, there exists a coded caching scheme for $(K_1, K_2, U,r, M, N)$ 2D caching-aided UDN system with the load
\begin{align*}
R_{\text{B}}=\left\{
	\begin{array}{lr}		
		\frac{3{K_1K_2-1 \choose t+1}+{K_1K_2-1 \choose t}+2{K_1K_2-3\choose t}}{{K_1K_2 \choose t}},\qquad \qquad \qquad \qquad \qquad \qquad \ \ \ \   \ \ \text{if } r=\frac{\sqrt{2}}{2},\\
		\frac{8{K_1K_2-1 \choose t+1}+\sum\limits_{i=1}^{7}{K_1K_2-i \choose t}-6{K_1K_2-2\choose t}+{K_1K_2-3 \choose t}+3{K_1K_2-4 \choose t}+4{K_1K_2-5 \choose t}}{{K_1K_2 \choose t}}, \quad\text{if } \frac{\sqrt{2}}{2}< r < 1,\\
		\frac{7{K_1K_2-1 \choose t+1}+\sum\limits_{i=2}^{7}{K_1K_2-i \choose t}-6{K_1K_2-2\choose t}+{K_1K_2-3 \choose t}+3{K_1K_2-4 \choose t}+4{K_1K_2-5 \choose t}}{{K_1K_2 \choose t}},  \quad\text{if } r=1.
	\end{array}
	\right.
\end{align*}
Moreover, if $K_1K_2\rightarrow \infty$, we have
\begin{align}\label{R1-approx}
R_{\text{B}}\approx\left\{
	\begin{array}{lr}		3\frac{N}{M}\left(1-\frac{M}{N}\right)^2+\left(1-\frac{M}{N}\right)+2\left(1-\frac{M}{N}\right)^3,\qquad \qquad \qquad\qquad \qquad\ \ \text{if } r=\frac{\sqrt{2}}{2},\\
	\left(8\frac{N}{M}-6\right)\left(1-\frac{M}{N}\right)^2+3\left(1-\frac{M}{N}\right)^4+4\left(1-\frac{M}{N}\right)^5+\sum\limits_{i=1}^{7}\left(1-\frac{M}{N}\right)^i, \ \ \text{if } \frac{\sqrt{2}}{2}< r < 1,\\
	\left(7\frac{N}{M}-6\right)\left(1-\frac{M}{N}\right)^2+3\left(1-\frac{M}{N}\right)^4+4\left(1-\frac{M}{N}\right)^5+\sum\limits_{i=2}^{7}\left(1-\frac{M}{N}\right)^i, \ \ \text{if } r=1.
	\end{array}
	\right.
\end{align}
\end{theorem}

\begin{remark}\rm Form \eqref{eq-load-baseline_all-approx} and \eqref{R1-approx}, we can obtain the ratio of the transmission load of Scheme B to the transmission load of Scheme A for the memory ratio of cache node $\frac{M}{N}$ when $K_1K_2\rightarrow \infty$,
\begin{align*}
	\frac{R_{\text{B}}}{R_{\text{A}}}\approx
	\left\{
	\begin{array}{lr}
	1+\frac{2M}{3N}\left(\left(1-\frac{M}{N}\right)^2-1\right),\qquad \qquad \qquad\qquad \qquad \qquad \qquad \qquad \qquad \ \ \text{if } r=\frac{\sqrt{2}}{2},\\
	\frac{M}{8N}\left(\left(\frac{8N}{M}-6\right)\left(1-\frac{M}{N}\right)+3\left(1-\frac{M}{N}\right)^3+4\left(1-\frac{M}{N}\right)^4+\sum\limits_{i=0}^{6}\left(1-\frac{M}{N}\right)^{i}\right),\ \text{if } \frac{\sqrt{2}}{2}< r < 1,\\
	\frac{M}{7N}\left(\left(\frac{7N}{M}-6\right)\left(1-\frac{M}{N}\right)+3\left(1-\frac{M}{N}\right)^3+4\left(1-\frac{M}{N}\right)^4+\sum\limits_{i=1}^{6}\left(1-\frac{M}{N}\right)^{i}\right), \ \text{if } r=1.
	\end{array}
	\right.
\end{align*}
\end{remark}

\subsection{Numerical Results}
\begin{figure}[t]
	\centering
	\subfigure[$\frac{\sqrt{2}}{2}< r < 1$, $K_1=K_2=6$, $U=288$ and $N=288$]{
		\includegraphics[scale=0.53]{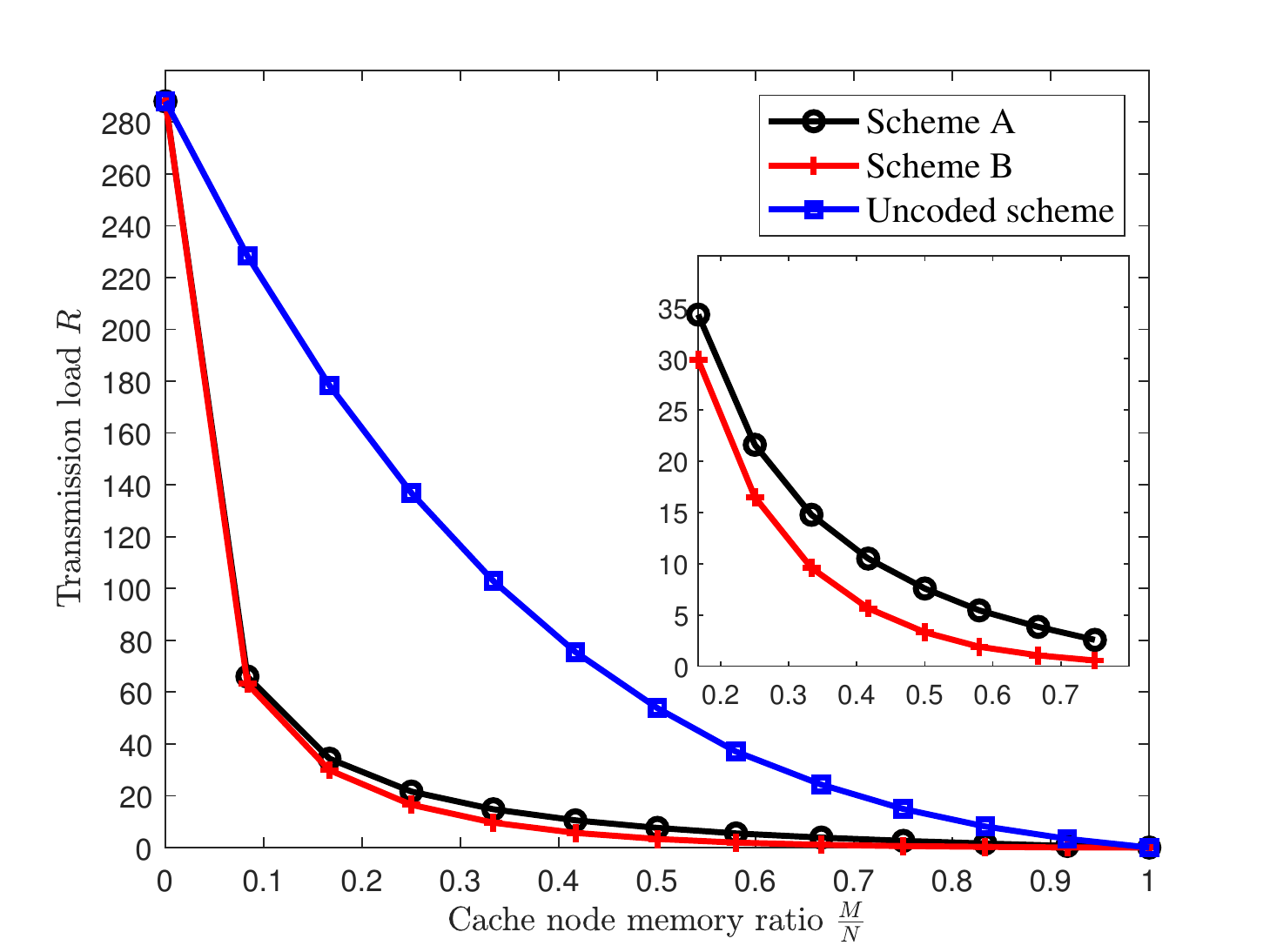}
		\label{fig:R-t_6x6}
	}
	\subfigure[$\frac{\sqrt{2}}{2}< r < 1$, $U=N=8K_1K_2$ and $\frac{M}{N}=\frac{1}{3}$]{
		\includegraphics[scale=0.53]{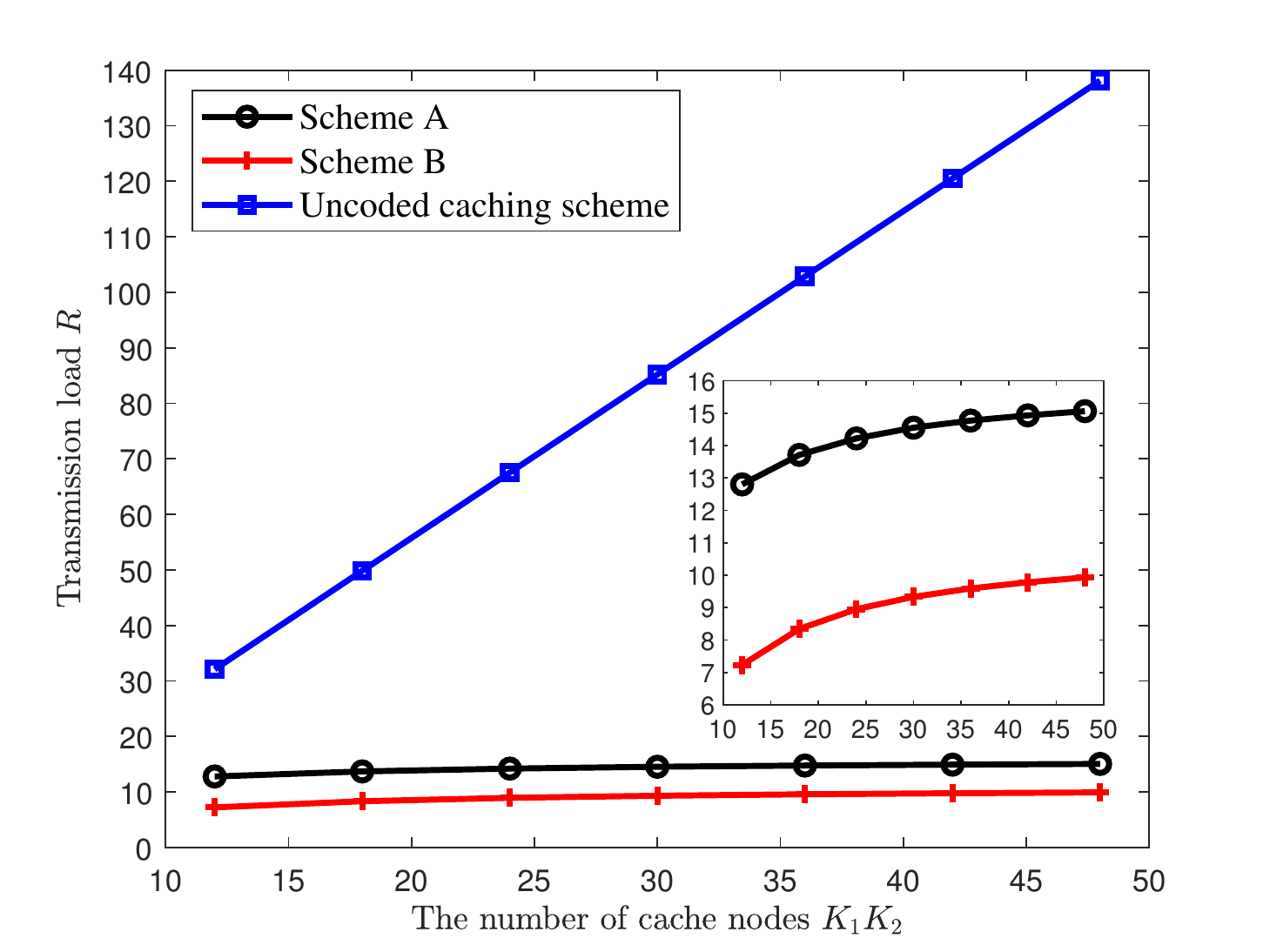}
		\label{fig:R-K1K2_1/3}
	}
	\caption{Comparison of  schemes for the $(K_1, K_2, U,r, M, N)$ 2D caching-aided UDN system.}
	\label{fig:3-R-t,K1K2_1}
\end{figure}

In this subsection, we compare the performance of Scheme A, Scheme B, and the conventional uncoded caching scheme Note that the transmission load of uncoded caching scheme for the $(K,M,N)$ shared-link broadcast system is $K\left(1-\frac{M}{N}\right)$ \cite{MN}. For the $(K_1, K_2, U,r, M, N)$ 2D caching-aided UDN system with $\frac{\sqrt{2}}{2} < r < 1$, the uncoded scheme is equivalent to using the uncoded caching scheme subsequently four shared-link broadcast
 systems respectively. Specifically, from \eqref{allusers} and \eqref{user_M/N}, each Type I user accesses a cache node and retrieves $M_{\mathcal{\text{I}}}$ files, it is equivalent to using a $(K_1K_2,M_{\mathcal{\text{I}}}, N)$ uncoded caching scheme with transmission load $R=K_1K_2(1-\lambda_1)$; Each Type II user accesses two cache nodes and retrieves $M_{\mathcal{\text{II}}}$ files, it is equivalent to using a $(2K_1K_2,M_{\mathcal{\text{II}}}, N)$ uncoded caching scheme with transmission load $R=2K_1K_2(1-\lambda_2)$; Each Type III user accesses three cache nodes and retrieves $M_{\mathcal{\text{II}}}$ files, it is equivalent to using a $(4K_1K_2,M_{\mathcal{\text{III}}}, N)$ uncoded caching scheme with transmission load $R=4K_1K_2(1-\lambda_3)$; Each Type IV user accesses four cache nodes and retrieves $M_{\mathcal{\text{IV}}}$ files, it is equivalent to using a $(K_1K_2,M_{\mathcal{\text{IV}}}, N)$ uncoded caching scheme with transmission load $R=K_1K_2(1-\lambda_4)$. Thus, the total transmission load of the conventional uncoded caching scheme is\vspace{-1ex}
\begin{align*}
	R_{\text{C}}=K_1K_2(1-\lambda_1)+2K_1K_2(1-\lambda_2)+4K_1K_2(1-\lambda_3)+K_1K_2(1-\lambda_4).
\end{align*}
 Fig. \ref{fig:R-t_6x6} compares the transmission load of the three schemes versus the cache node memory ratio $M/N$ when  $\frac{\sqrt{2}}{2} < r < 1$, $K_1=K_2=6$, $U=288$ and $N=288$. It can be seen that the transmission load of the uncoded caching scheme is much greater than that of Scheme A and Scheme B. In addition, we can also see that the transmission load of Scheme B has a certain extent reduction based on Scheme A. In Fig. \ref{fig:R-K1K2_1/3}, we compare the transmission load of the three schemes when the number of cache nodes increases when $\frac{\sqrt{2}}{2}< r < 1$ and $\frac{M}{N}=\frac{1}{3}$. Notice that the increase of cache nodes will lead to the increase of files and users in our hypothetical 2D caching-aided UDN system, i.e., the number of users increases with the number of cache nodes in Fig. \ref{fig:R-K1K2_1/3}. It can be seen that the transmission load of the conventional uncoded caching scheme linearly increases the number of $K_1K_2$, while our proposed schemes demonstrate more robust behavior against the variation of $K_1K_2$. Besides, there is a clear gap between Scheme B and Scheme A, and both Scheme A and Scheme B are approaching constant values when $K_1K_2$ is sufficiently large.

\subsection{Example For Theorem \ref{thm-scheme1}}
In this subsection, we also use Example \ref{e.g-users} with the parameters $K_1=K_2=3$, $K_1K_2=9$, $M=16$, $N=72$ and $U=72$ to further explain the redundant transmitted coded signals and the modified transmitted coded signals of our Scheme B. Clearly, we have $t=\frac{K_1K_2M}{N}=2$. We can obtain a $(3,3,72,r,16,72)$ 2D caching-aided UDN scheme by the $(9,16,72)$ MN scheme and three MDS codes where $\frac{\sqrt{2}}{2}<r<1$.

\textbf{Placement phase:} By Lines 1-5 of Algorithm \ref{alg-MN}, each file $W_n$ where $n \in [0:9)$ is divided into ${9 \choose 2}=36$ packets, and each packet is labeled by $2$-subset $\mathcal{T}\in {\mathcal{K} \choose 2}$ where $\mathcal{K}=[0:3)\times [0:3)$, then from \eqref{SC1-Z_Ck} each cache node directly caches $\frac{M}{N}F$= $\frac{16}{72} \times 36$ = $8$ packets of each file. Each user can retrieve packets from its accessible cache node(s) from \eqref{SC1-Z_Uk}. Here we only list the details of the Type II-1 user $\mathcal{U}^{\text{II-1}}_{0,0}=\{(0,0),(0,1)\}$, Type III-1 user $\mathcal{U}^{\text{III-1}}_{0,0}=\{(0,0),(0,1),(1,0)\}$ and Type IV user $\mathcal{U}^{\text{IV}}_{0,0}=\{(0,0),(0,1),(1,0),(1,1)\}$, because other users in each type are similar. From \eqref{SC1-Z_Uk}, users $\mathcal{U}^{\text{II-1}}_{0,0}$, $\mathcal{U}^{\text{III-1}}_{0,0}$ and $\mathcal{U}^{\text{IV}}_{0,0}$ can obtain the following packets respectively.
	\begin{align}
\mathcal{Z}_{\mathcal{U}^{\text{II-1}}_{0,0}}=\{&W_{n,\{(0,0),(0,1)\}},W_{n,\{(0,0),(0,2)\}},W_{n,\{(0,0),(1,0)\}},W_{n,\{(0,0),(1,1)\}},W_{n,\{(0,0),(1,2)\}},W_{n,\{(0,0),(2,0)\}},\nonumber\\
&W_{n,\{(0,0),(2,1)\}},W_{n,\{(0,0),(2,2)\}},W_{n,\{(0,1),(0,2)\}},W_{n,\{(0,1),(1,0)\}},W_{n,\{(0,1),(1,1)\}},W_{n,\{(0,1),(1,2)\}},\nonumber\\
&W_{n,\{(0,1),(2,0)\}},W_{n,\{(0,1),(2,1)\}},W_{n,\{(0,1),(2,2)\}}\}\ \text{where } n \in [0:9),\label{eq-packets-II}
	\end{align}
	\begin{align}
\mathcal{Z}_{\mathcal{U}^{\text{III-1}}_{0,0}}=\{&W_{n,\{(0,0),(0,1)\}},W_{n,\{(0,0),(0,2)\}},W_{n,\{(0,0),(1,0)\}},W_{n,\{(0,0),(1,1)\}},W_{n,\{(0,0),(1,2)\}},W_{n,\{(0,0),(2,0)\}},\nonumber\\
&W_{n,\{(0,0),(2,1)\}},W_{n,\{(0,0),(2,2)\}},W_{n,\{(0,1),(0,2)\}},W_{n,\{(0,1),(1,0)\}},W_{n,\{(0,1),(1,1)\}},W_{n,\{(0,1),(1,2)\}},\nonumber\\
&W_{n,\{(0,1),(2,0)\}},W_{n,\{(0,1),(2,1)\}},W_{n,\{(0,1),(2,2)\}},W_{n,\{(0,2),(1,0)\}},W_{n,\{(1,0),(1,1)\}},W_{n,\{(1,0),(1,2)\}},\nonumber\\
&W_{n,\{(1,0),(2,0)\}},W_{n,\{(1,0),(2,1)\}},W_{n,\{(1,0),(2,2)\}}\}\ \text{where } n \in [0:9),\label{eq-packets-III}
	\end{align}
	\begin{align}
\mathcal{Z}_{\mathcal{U}^{\text{IV}}_{0,0}}=\{&W_{n,\{(0,0),(0,1)\}},W_{n,\{(0,0),(0,2)\}},W_{n,\{(0,0),(1,0)\}},W_{n,\{(0,0),(1,1)\}},W_{n,\{(0,0),(1,2)\}},W_{n,\{(0,0),(2,0)\}},\nonumber\\
&W_{n,\{(0,0),(2,1)\}},W_{n,\{(0,0),(2,2)\}},W_{n,\{(0,1),(0,2)\}},W_{n,\{(0,1),(1,0)\}},W_{n,\{(0,1),(1,1)\}},W_{n,\{(0,1),(1,2)\}},\nonumber\\
&W_{n,\{(0,1),(2,0)\}},W_{n,\{(0,1),(2,1)\}},W_{n,\{(0,1),(2,2)\}},W_{n,\{(0,2),(1,0)\}},W_{n,\{(1,0),(1,1)\}},W_{n,\{(1,0),(1,2)\}},\nonumber\\
&W_{n,\{(1,0),(2,0)\}},W_{n,\{(1,0),(2,1)\}},W_{n,\{(1,0),(2,2)\}},W_{n,\{(0,2),(1,1)\}},W_{n,\{(1,1),(1,2)\}},W_{n,\{(1,1),(2,0)\}},\nonumber\\
&W_{n,\{(1,1),(2,1)\}},W_{n,\{(1,1),(2,2)\}}\}\ \text{where } n \in [0:9).\label{eq-packets-IV}
	\end{align}
\textbf{Delivery phase}: Assume that each user randomly request one file. According to the Lines 10-12 of Algorithm \ref{alg-MN}, the server sends $S={9\choose 3}=84$ coded signals $X_{\mathcal{S}}$ where $\mathcal{S}\in {\mathcal{K}\choose 3}$ for Type I user. Let us see user $\mathcal{U}^{\text{II-1}}_{0,0}$, $\mathcal{U}^{\text{III-1}}_{0,0}$ and $\mathcal{U}^{\text{IV}}_{0,0}$ to show the redundancy in Scheme A.

Let us introduce the $h^{\text{II}}_1$ and $h^{\text{II}}_2$ coded (and modified coded) signals for user $\mathcal{U}^{\text{II-1}}_{0,0}$ of Type II-1.
\begin{itemize}
	\item Firstly, we count the value of $h^{\text{II}}_1$ and find the $h^{\text{II}}_1$ coded signals. We can obtain the following $2$-subsets $\mathcal{T}_1$,
	\begin{align*}
		&\{(0,1),(0,2)\},\ \{(0,1),(1,0)\},\ \{(0,1),(1,1)\}, \{(0,1),(1,2)\},\\
		&\{(0,1),(2,0)\},\ \{(0,1),(2,1)\},\ \{(0,1),(2,2)\},
	\end{align*}
	satisfying condition \eqref{cond:II-1_T1}, i.e., $(0,0)\notin \mathcal{T}_1$ and $(0,1)\in \mathcal{T}_1$. So we have
	$h^{\text{II}}_1={K_1K_2-1 \choose t}-{K_1K_2-2\choose t}={9-1\choose 2}-{9-2\choose 2}=28-21=7$.
	Let us first consider $\mathcal{T}_1=\{(0,1),(0,2)\}$. From \eqref{eq-packets-II}, the packets in coded signal
	\begin{align*}
		X_{\{(0,0)\}\bigcup\mathcal{T}_1 }=X_{\{(0,0),(0,1),(0,2)\}}=W_{d_{\mathcal{U}^{\text{II-1}}_{0,0}},\{(0,1),(0,2)\}}\bigoplus W_{d_{\mathcal{U}^{\text{II-1}}_{0,1}},\{(0,0),(0,2)\}}\bigoplus W_{d_{\mathcal{U}^{\text{II-1}}_{0,2}},\{(0,0),(0,1)\}}
	\end{align*}
	can be retrieved by user $\mathcal{U}^{\text{II-1}}_{0,0}$ through its connected cache nodes $\text{C}_{0,0}$ and $\text{C}_{0,1}$. Similarly we can check that user $\mathcal{U}^{\text{II-1}}_{0,0}$ can retrieve all the packets in the following coded signals
	\begin{align}
		&X_{\{(0,0),(0,1),(0,2)\}}, X_{\{(0,0),(0,1),(1,0)\}}, X_{\{(0,0),(0,1),(1,1)\}}, X_{\{(0,0),(0,1),(1,2)\}},\nonumber\\
		& X_{\{(0,0),(0,1),(2,0)\}},X_{\{(0,0),(0,1),(2,1)\}}, X_{\{(0,0),(0,1),(2,2)\}}.\label{eq:II-1-cancelsignal-S1}
	\end{align}
		
\item Now let us count the value of $h^{\text{II}}_2$ and find the $h^{\text{II}}_2$ (modified) coded signals. We can obtain all the following $2$-subsets $\mathcal{T}_2$,
\begin{align*}
	\{(0,2),(1,0)\},\{(0,2),(1,1)\},\{(0,2),(1,2)\},\{(0,2),(2,0)\},\{(0,2),(2,1)\},\{(0,2),(2,2)\},
\end{align*}
satisfying condition \eqref{cond:II-1_T2}, i.e., $(0,0)\notin \mathcal{T}_2$, $(0,1)\notin \mathcal{T}_2$ and $(0,2)\in \mathcal{T}_2$. So we have
$h^{\text{II}}_2={K_1K_2-2 \choose t}-{K_1K_2-3\choose t}={9-2 \choose 2}-{9-3\choose 2}=21-15=6$.
Let us first consider $\mathcal{T}_2=\{(0,2),(1,0)\}$. The packets in coded signal
$$
X_{\{(0,1)\}\bigcup \mathcal{T}_2}=X_{\{(0,1),(0,2),(1,0)\}}=W_{d_{\mathcal{U}^{\text{II-1}}_{0,1}},\{(0,2),(1,0)\}}\bigoplus W_{d_{\mathcal{U}^{\text{II-1}}_{0,2}},\{(0,1),(1,0)\}}\bigoplus W_{d_{\mathcal{U}^{\text{II-1}}_{1,0}},\{(0,1),(0,2)\}}
$$
can be retrieved by user $\mathcal{U}^{\text{II-1}}_{0,1}$ through its connected cache nodes $\text{C}_{0,1}$ and $\text{C}_{0,2}$, then the server only needs to send the following modified coded signal
$$X'_{\{(0,1),(0,2),(1,0)\}}=X_{\{(0,1),(0,2),(1,0)\}}-W_{d_{\mathcal{U}^{\text{II-1}}_{0,1}},\{(0,2),(1,0)\}}=W_{d_{\mathcal{U}^{\text{II-1}}_{0,2}},\{(0,1),(1,0)\}}\bigoplus W_{d_{\mathcal{U}^{\text{II-1}}_{1,0}},\{(0,1),(0,2)\}}.$$
Furthermore from \eqref{eq-packets-II}, user $\mathcal{U}^{\text{II-1}}_{0,0}$ can retrieve all the packets in $X'_{\{(0,1),(0,2),(1,0)\}}$ from its connected cache node $\text{C}_{0,1}$. Similarly we can check that user $\mathcal{U}^{\text{II-1}}_{0,0}$ can retrieve all the packets in modified coded signals
\begin{align*}
	&X'_{\{(0,1),(0,2),(1,0)\}},\ X'_{\{(0,1),(0,2),(1,1)\}},\ X'_{\{(0,1),(0,2),(1,2)\}},\nonumber\\
	&X'_{\{(0,1),(0,2),(2,0)\}},\ X'_{\{(0,1),(0,2),(2,1)\}},\ X'_{\{(0,1),(0,2),(2,2)\}},
\end{align*}
which are generated by the following coded signals respectively
\begin{align}
	\label{eq:II-1-cancelsignal-S2}
	&X_{\{(0,1),(0,2),(1,0)\}},\ X_{\{(0,1),(0,2),(1,1)\}},\ X_{\{(0,1),(0,2),(1,2)\}},\nonumber\\ &X_{\{(0,1),(0,2),(2,0)\}},\ X_{\{(0,1),(0,2),(2,1)\}},\ X_{\{(0,1),(0,2),(2,2)\}}.
\end{align}
\end{itemize}
	Obviously, the $h^{\text{II}}_1$ coded signals in \eqref{eq:II-1-cancelsignal-S1} are different from the $h^{\text{II}}_2$ coded signals in \eqref{eq:II-1-cancelsignal-S2}. So there are exactly $h^{\text{II}}=h^{\text{II}}_{1}+h^{\text{II}}_{2}=13$ coded (or modified coded) signals which can be retrieved by user $\mathcal{U}^{\text{II-1}}_{0,0}$.

Let us introduce the $h^{\text{III}}_1$, $h^{\text{III}}_2$ and $h^{\text{III}}_3$ coded (and modified coded) signals for user $\mathcal{U}^{\text{III-1}}_{0,0}$ of III-1.
	\begin{itemize}
		\item Firstly, we count the value of $h^{\text{III}}_1$ and find the $h^{\text{III}}_1$ coded signals. We can obtain the following $2$-subsets $\mathcal{T}_1$,
		\begin{align*}
			&\{(0,1),(0,2)\},\ \{(0,1),(1,0)\},\ \{(0,1),(1,1)\},\ \{(0,1),(1,2)\},\ \{(0,1),(2,0)\}, \\
			&\{(0,1),(2,1)\},\ \{(0,1),(2,2)\},\ \{(0,2),(1,0)\},\ \{(1,0),(1,1)\},\ \{(1,0),(1,2)\}, \\
			&\{(1,0),(2,0)\},\ \{(1,0),(2,1)\},\ \{(1,0),(2,2)\},
		\end{align*}
		satisfying condition \eqref{cond:III-1_T1}, i.e., $(0,0)\notin \mathcal{T}_1$ and $\{(0,1),(1,0)\}\bigcap \mathcal{T}_1\neq \emptyset$. So we have
		$
		h^{\text{III}}_1={K_1K_2-1 \choose t}-{K_1K_2-3\choose t}={9-1\choose 2}-{9-3\choose 2}=28-15=13$.	
		Let us first consider $\mathcal{T}_1=\{(0,1),(0,2)\}$. From \eqref{eq-packets-III}, the packets in coded signal
		$$
		X_{\{(0,0)\}\bigcup \mathcal{T}_1}=X_{\{(0,0),(0,1),(0,2)\}}=W_{d_{\mathcal{U}^{\text{III-1}}_{0,0}},\{(0,1),(0,2)\}}\bigoplus W_{d_{\mathcal{U}^{\text{III-1}}_{0,1}},\{(0,0),(0,2)\}}\bigoplus W_{d_{\mathcal{U}^{\text{III-1}}_{0,2}},\{(0,0),(0,1)\}}
		$$
		can be retrieved by user $\mathcal{U}^{\text{III-1}}_{0,0}$ through its connected cache nodes $\text{C}_{0,0}$ and $\text{C}_{0,1}$. Similarly we can check that user $\mathcal{U}^{\text{III-1}}_{0,0}$ can also retrieve all the packets in the following coded signals
		\begin{align}
			&X_{\{(0,0),(0,1),(0,2)\}}, X_{\{(0,0),(0,1),(1,0)\}}, X_{\{(0,0),(0,1),(1,1)\}}, X_{\{(0,0),(0,1),(1,2)\}}, X_{\{(0,0),(0,1),(2,0)\}}, \nonumber\\
			&X_{\{(0,0),(0,1),(2,1)\}}, X_{\{(0,0),(0,1),(2,2)\}}, X_{\{(0,0),(0,2),(1,0)\}}, X_{\{(0,0),(1,0),(1,1)\}}, X_{\{(0,0),(1,0),(1,2)\}}, \nonumber\\
			&X_{\{(0,0),(1,0),(2,0)\}}, X_{\{(0,0),(1,0),(2,1)\}}, X_{\{(0,0),(1,0),(2,2)\}}.\label{eq:III-1-cancelsignal-S1}
		\end{align}
		
		\item Then we count the value of $h^{\text{III}}_2$ and find the $h^{\text{III}}_2$ (modified) coded signals. We can obtain the following $2$-subsets $\mathcal{T}_2$,
		\begin{align*}
			&\{(0,2),(1,0)\},\ \{(0,2),(1,1)\},\ \{(0,2),(1,2)\},\ \{(0,2),(2,0)\},\\
			&\{(0,2),(2,1)\},\ \{(0,2),(2,2)\},\ \{(1,1),(1,2)\},\ \{(1,1),(1,2)\},\\
			&\{(1,1),(2,0)\},\ \{(1,1),(2,1)\},\ \{(1,1),(2,2)\},
		\end{align*}
		satisfying condition \eqref{cond:III-1_T2}, i.e., $(0,0)\notin \mathcal{T}_2$, $(0,1)\notin \mathcal{T}_2$ and $\{(0,2),(1,1)\}\bigcap \mathcal{T}_2\neq \emptyset$. So we have
		$h^{\text{III}}_2={K_1K_2-2 \choose t}-{K_1K_2-4\choose t}={9-2\choose 2}-{9-4\choose 2}=21-10=11$.
		Let us first consider $\mathcal{T}_2=\{(0,2),(1,0)\}$. The packets in coded signal
		$$
		X_{\{(0,1)\}\bigcup \mathcal{T}_2}=X_{\{(0,1),(0,2),(1,0)\}}=W_{d_{\mathcal{U}^{\text{III-1}}_{0,1}},\{(0,2),(1,0)\}}\bigoplus W_{d_{\mathcal{U}^{\text{III-1}}_{0,2}},\{(0,1),(1,0)\}}\bigoplus W_{d_{\mathcal{U}^{\text{III-1}}_{1,0}},\{(0,1),(0,2)\}}
		$$
		can be retrieved by user $\mathcal{U}^{\text{III-1}}_{0,1}$ through its connected cache nodes $\text{C}_{0,1}$ and $\text{C}_{0,2}$, then the server only needs to send the following modified coded signal
		$$X'_{\{(0,1),(0,2),(1,0)\}}=X_{\{(0,1),(0,2),(1,0)\}}-W_{d_{\mathcal{U}^{\text{III-1}}_{0,1}},\{(0,2),(1,0)\}}=W_{d_{\mathcal{U}^{\text{III-1}}_{0,2}},\{(0,1),(1,0)\}}\bigoplus W_{d_{\mathcal{U}^{\text{III-1}}_{1,0}},\{(0,1),(0,2)\}}.$$
		From \eqref{eq-packets-III}, user $\mathcal{U}^{\text{III-1}}_{0,0}$ can retrieve all the packets in $X'_{\{(0,1),(0,2),(1,0)\}}$ from the its connected cache node $\text{C}_{0,1}$. Similarly we can check that user $\mathcal{U}^{\text{III-1}}_{0,0}$ can also retrieve all the packets in the following modified coded signals
		\begin{align*}
			&X'_{\{(0,1),(0,2),(1,0)\}},\ X'_{\{(0,1),(0,2),(1,1)\}},\ X'_{\{(0,1),(0,2),(1,2)\}},\ X'_{\{(0,1),(0,2),(2,0)\}},\\
			&X'_{\{(0,1),(0,2),(2,1)\}},\ X'_{\{(0,1),(0,2),(2,2)\}},\
			X'_{\{(0,1),(1,1),(1,2)\}},\ X'_{\{(0,1),(1,1),(1,2)\}},\\
			&X'_{\{(0,1),(1,1),(2,0)\}},\ X'_{\{(0,1),(1,1),(2,1)\}},\ X'_{\{(0,1),(1,1),(2,2)\}},
		\end{align*}
		which are generated by the following coded signals respectively
		\begin{align}
			&X_{\{(0,1),(0,2),(1,0)\}},\ X_{\{(0,1),(0,2),(1,1)\}},\ X_{\{(0,1),(0,2),(1,2)\}},\ X_{\{(0,1),(0,2),(2,0)\}},\nonumber\\
			&X_{\{(0,1),(0,2),(2,1)\}},\ X_{\{(0,1),(0,2),(2,2)\}},\
			X_{\{(0,1),(1,1),(1,2)\}},\ X_{\{(0,1),(1,1),(1,2)\}},\nonumber\\
			&X_{\{(0,1),(1,1),(2,0)\}},\ X_{\{(0,1),(1,1),(2,1)\}},\ X_{\{(0,1),(1,1),(2,2)\}}.\label{eq:III-1-cancelsignal-S2}
		\end{align}
		
		\item We count the value of $h^{\text{III}}_3$ and find the $h^{\text{III}}_3$ coded signals. We can obtain the following $2$-subsets $\mathcal{T}_3$,
		\begin{align*}
			&\{(0,2),(1,1)\},\ \{(1,1),(1,2)\},\ \{(1,1),(2,0)\},\ \{(1,1),(2,1)\},\ \{(1,1),(2,2)\}, \\
			&\{(0,2),(2,0)\},\ \{(1,2),(2,0)\},\ \{(2,0),(2,1)\},\ \{(2,0),(2,2)\}.
		\end{align*}
		satisfying condition \eqref{cond:III-1_T3}, i.e., $(0,0)\notin \mathcal{T}_3$, $(0,1)\notin \mathcal{T}_2$, $(1,0)\notin \mathcal{T}_3$ and $\{(1,1),(2,0)\}\bigcap \mathcal{T}_3\neq \emptyset$. So we have
		$h^{\text{III}}_3={K_1K_2-3 \choose t}-{K_1K_2-5\choose t}={9-3\choose 2}-{9-5\choose 2}=15-6=9$.
		Let us first consider $\mathcal{T}_3=\{(0,2),(1,1)\}$. The packets in coded signal
		$$
		X_{\{(1,0)\}\bigcup \mathcal{T}_3}=X_{\{(0,2),(1,0),(1,1)\}}=W_{d_{\mathcal{U}^{\text{III-1}}_{0,2}},\{(1,0),(1,1)\}}\bigoplus W_{d_{\mathcal{U}^{\text{III-1}}_{1,0}},\{(0,2),(1,1)\}}\bigoplus W_{d_{\mathcal{U}^{\text{III-1}}_{1,1}},\{(0,2),(1,0)\}}
		$$
		can be retrieved by user $\mathcal{U}^{\text{III-1}}_{1,0}$ through its connected cache nodes $\text{C}_{1,0}$ and $\text{C}_{1,1}$, then the server only needs to send the following modified coded signal
		$$
		X'_{\{(0,2),(1,0),(1,1)\}}=X_{\{(0,2),(1,0),(1,1)\}}-W_{d_{\mathcal{U}^{\text{III-1}}_{1,0}},\{(0,2),(1,1)\}}=W_{d_{\mathcal{U}^{\text{III-1}}_{0,2}},\{(1,0),(1,1)\}}\bigoplus W_{d_{\mathcal{U}^{\text{III-1}}_{1,1}},\{(0,2),(1,0)\}}.
		$$
		From \eqref{eq-packets-III}, user $\mathcal{U}^{\text{III-1}}_{0,0}$ can retrieve all the packets in $X'_{\{(0,2),(1,0),(1,1)\}}$ from its connected cache node $\text{C}_{1,0}$. Similarly we can check that user $\mathcal{U}^{\text{III-1}}_{0,0}$ can retrieve all the packets in the following modified coded signals
		\begin{align*}
			&X'_{\{(0,2),(1,0),(1,1)\}},\ X'_{\{(0,2),(1,0),(2,0)\}},\ X'_{\{(1,0),(1,1),(1,2)\}},\ X'_{\{(1,0),(1,1),(2,0)\}},\ X'_{\{(1,0),(1,1),(2,1)\}}, \\
			&X'_{\{(1,0),(1,1),(2,2)\}},\ X'_{\{(1,0),(1,2),(2,0)\}},\ X'_{\{(1,0),(2,0),(2,1)\}},\ X'_{\{(1,0),(2,0),(2,2)\}},
		\end{align*}
		which are generated by the following coded signals respectively
		\begin{align}
			&X_{\{(0,2),(1,0),(1,1)\}},\ X_{\{(0,2),(1,0),(2,0)\}},\ X_{\{(1,0),(1,1),(1,2)\}},\ X_{\{(1,0),(1,1),(2,0)\}},\ X_{\{(1,0),(1,1),(2,1)\}}, \nonumber\\
			&X_{\{(1,0),(1,1),(2,2)\}},\ X_{\{(1,0),(1,2),(2,0)\}},\ X_{\{(1,0),(2,0),(2,1)\}},\ X_{\{(1,0),(2,0),(2,2)\}}.\label{eq:III-1-cancelsignal-S3}
		\end{align}
	\end{itemize}
	Obviously, the $h^{\text{III}}_1$, $h^{\text{III}}_2$ and $h^{\text{III}}_3$ coded signals in \eqref{eq:III-1-cancelsignal-S1}, \eqref{eq:III-1-cancelsignal-S2} and \eqref{eq:III-1-cancelsignal-S3} are different from each other. So there are exactly $h^{\text{III}}=h^{\text{III}}_{1}+h^{\text{III}}_{2}+h^{\text{III}}_{3}=33$ coded (or modified coded) signals which can be retrieved by user $\mathcal{U}^{\text{III-1}}_{0,0}$.
	

Let us introduce the $h^{\text{IV}}_1$, $h^{\text{IV}}_2$, $h^{\text{IV}}_3$ and $h^{\text{IV}}_4$ coded (and modified coded) signals for user $\mathcal{U}^{\text{IV-1}}_{0,0}$ of Type IV.
\begin{itemize}
	\item Firstly, we count the value of $h^{\text{IV}}_1$ and find the $h^{\text{IV}}_1$ coded signals. We can obtain the following $2$-subsets $\mathcal{T}_1$,
	\begin{align*}
		&\{(0,1),(0,2)\},\ \{(0,1),(1,0)\},\ \{(0,1),(1,1)\},\ \{(0,1),(1,2)\},\ \{(0,1),(2,0)\},\\
		&\{(0,1),(2,1)\},\ \{(0,1),(2,2)\},\ \{(0,2),(1,0)\},\ \{(0,2),(1,1)\},\ \{(1,0),(1,1)\},\\
		&\{(1,0),(1,2)\},\ \{(1,0),(2,0)\},\ \{(1,0),(2,1)\},\ \{(1,0),(2,2)\},\ \{(1,1),(1,2)\},\\
		&\{(1,1),(2,0)\},\ \{(1,1),(2,1)\},\ \{(1,1),(2,2)\},
	\end{align*}
	satisfying condition \eqref{cond:IV_T1}, i.e., $(0,0)\notin \mathcal{T}_1$ and $\{(0,1),(1,0),(1,1)\}\bigcap \mathcal{T}_1\neq \emptyset$. So we have
	$h^{\text{IV}}_1={K_1K_2-1 \choose t}-{K_1K_2-4\choose t}={9-1\choose 2}-{9-4\choose 2}=28-10=18$.
	Let us first consider $\mathcal{T}_1=\{(0,1),(0,2)\}$. From \eqref{eq-packets-IV}, the packets in coded signal
	$$
	X_{\{(0,0)\}\bigcup \mathcal{T}_1}=X_{\{(0,0),(0,1),(0,2)\}}=W_{d_{\mathcal{U}^{\text{IV}}_{0,0}},\{(0,1),(0,2)\}}\bigoplus W_{d_{\mathcal{U}^{\text{IV}}_{0,1}},\{(0,0),(0,2)\}}\bigoplus W_{d_{\mathcal{U}^{\text{IV}}_{0,2}},\{(0,0),(0,1)\}}
	$$
	can be retrieved by user $\mathcal{U}^{\text{IV}}_{0,0}$ through its connected cache nodes $\text{C}_{0,0}$ and $\text{C}_{0,1}$. Similarly we can check that user $\mathcal{U}^{\text{IV}}_{0,0}$ can retrieve all the packets in the following coded signals
	\begin{align}
		&X_{\{(0,0),(0,1),(0,2)\}},\ X_{\{(0,0),(0,1),(1,0)\}},\ X_{\{(0,0),(0,1),(1,1)\}},\ X_{\{(0,0),(0,1),(1,2)\}},\ X_{\{(0,0),(0,1),(2,0)\}},\nonumber\\ &X_{\{(0,0),(0,1),(2,1)\}},\ X_{\{(0,0),(0,1),(2,2)\}},\ X_{\{(0,0),(0,2),(1,0)\}},\ X_{\{(0,0),(0,2),(1,1)\}},\ X_{\{(0,0),(1,0),(1,1)\}},\nonumber\\
		&X_{\{(0,0),(1,0),(1,2)\}},\ X_{\{(0,0),(1,0),(2,0)\}},\ X_{\{(0,0),(1,0),(2,1)\}},\ X_{\{(0,0),(1,0),(2,2)\}},\ X_{\{(0,0),(1,1),(1,2)\}},\nonumber\\
		&X_{\{(0,0),(1,1),(2,0)\}},\ X_{\{(0,0),(1,1),(2,1)\}},\ X_{\{(0,0),(1,1),(2,2)\}}. \label{eq:IV-cancelsignal-S1}
	\end{align}
	
	\item Then we count the value of $h^{\text{IV}}_2$ and find the $h^{\text{IV}}_2$ coded signals. We can obtain the following $2$-subsets $\mathcal{T}_2$,
	\begin{align*}
		&\{(0,2),(1,0)\},\ \{(0,2),(1,1)\},\ \{(0,2),(1,2)\},\ \{(0,2),(2,0)\},\ \{(0,2),(2,1)\}, \\
		&\{(0,2),(2,2)\},\ \{(1,0),(1,1)\},\ \{(1,0),(1,2)\},\ \{(1,1),(1,2)\},\ \{(1,1),(2,0)\}, \\
		&\{(1,1),(2,1)\},\ \{(1,1),(2,2)\},\ \{(1,2),(2,0)\},\ \{(1,2),(2,1)\},\ \{(1,2),(2,2)\},
	\end{align*}
	satisfying condition \eqref{cond:IV_T2}, i.e., $(0,0)\notin \mathcal{T}_2$, $(0,1)\notin \mathcal{T}_2$ and $\{(0,2),(1,1),(1,2)\}\bigcap \mathcal{T}_2\neq \emptyset$. So we have
	$h^{\text{IV}}_2={K_1K_2-2 \choose t}-{K_1K_2-5\choose t}={9-2\choose 2}-{9-5\choose 2}=21-6=15$.
	Let us first consider $\mathcal{T}_2=\{(0,2),(1,0)\}$. The packets in coded signal
	$$
	X_{\{(0,1)\}\bigcup \mathcal{T}_2}=X_{\{(0,1),(0,2),(1,0)\}}=W_{d_{\mathcal{U}^{\text{IV}}_{0,1}},\{(0,2),(1,0)\}}\bigoplus W_{d_{\mathcal{U}^{\text{IV}}_{0,2}},\{(0,1),(1,0)\}}\bigoplus W_{d_{\mathcal{U}^{\text{IV}}_{1,0}},\{(0,1),(0,2)\}}
	$$
	can retrieved by user $\mathcal{U}^{\text{IV}}_{0,1}$ through its connected cache nodes $\text{C}_{0,1}$ and $\text{C}_{0,2}$, then the server only needs to send the following modified coded signal
	$$
	X'_{\{(0,1),(0,2),(1,0)\}}=X_{\{(0,1),(0,2),(1,0)\}}-W_{d_{\mathcal{U}^{\text{IV}}_{0,1}},\{(0,2),(1,0)\}}=W_{d_{\mathcal{U}^{\text{IV}}_{0,2}},\{(0,1),(1,0)\}}\oplus W_{d_{\mathcal{U}^{\text{IV}}_{1,0}},\{(0,1),(0,2)\}}.
	$$
	From \eqref{eq-packets-IV}, user $\mathcal{U}^{\text{IV}}_{0,0}$ can retrieve all the packets in $X'_{\{(0,1),(0,2),(1,0)\}}$ from its connected cache node $\text{C}_{0,1}$. Similarly we can check that user $\mathcal{U}^{\text{IV}}_{0,0}$ can retrieve all the packets in the following modified coded signals
	\begin{align*}
		&X'_{\{(0,1),(0,2),(1,0)\}},\ X'_{\{(0,1),(0,2),(1,1)\}},\ X'_{\{(0,1),(0,2),(1,2)\}},\ X'_{\{(0,1),(0,2),(2,0)\}},\ X'_{\{(0,1),(0,2),(2,1)\}},\\
		&X'_{\{(0,1),(0,2),(2,2)\}},\ X'_{\{(0,1),(1,0),(1,1)\}},\ X'_{\{(0,1),(1,0),(1,2)\}},\ X'_{\{(0,1),(1,1),(1,2)\}},\ X'_{\{(0,1),(1,1),(2,0)\}},\\
		&X'_{\{(0,1),(1,1),(2,1)\}},\ X'_{\{(0,1),(1,1),(2,2)\}},\ X'_{\{(0,1),(1,2),(2,0)\}},\ X'_{\{(0,1),(1,2),(2,1)\}},\ X'_{\{(0,1),(1,2),(2,2)\}},
	\end{align*}
	which are generated by the following coded signals respectively
	\begin{align}
		&X_{\{(0,1),(0,2),(1,0)\}},\ X_{\{(0,1),(0,2),(1,1)\}},\ X_{\{(0,1),(0,2),(1,2)\}},\ X_{\{(0,1),(0,2),(2,0)\}},\ X_{\{(0,1),(0,2),(2,1)\}},\nonumber\\
		&X_{\{(0,1),(0,2),(2,2)\}},\ X_{\{(0,1),(1,0),(1,1)\}},\ X_{\{(0,1),(1,0),(1,2)\}},\ X_{\{(0,1),(1,1),(1,2)\}},\ X_{\{(0,1),(1,1),(2,0)\}},\nonumber\\
		&X_{\{(0,1),(1,1),(2,1)\}},\ X_{\{(0,1),(1,1),(2,2)\}},\ X_{\{(0,1),(1,2),(2,0)\}},\ X_{\{(0,1),(1,2),(2,1)\}},\ X_{\{(0,1),(1,2),(2,2)\}}.\label{eq:IV-cancelsignal-S2}
	\end{align}
	
	\item Next, we count the value of $h^{\text{IV}}_3$ and find the $h^{\text{IV}}_3$ coded signals. We can get the following $2$-subsets $\mathcal{T}_3$,
	\begin{align*}
		&\{(0,2),(1,1)\},\ \{(0,2),(2,0)\},\ \{(0,2),(2,1)\},\ \{(1,1),(1,2)\},\\
		&\{(1,1),(2,0)\},\ \{(1,1),(2,1)\},\ \{(1,1),(2,2)\},\ \{(1,2),(2,0)\},\\
		&\{(1,2),(2,1)\},\ \{(2,0),(2,1)\},\ \{(2,0),(2,2)\},\ \{(2,1),(2,2)\},
	\end{align*}
	satisfying condition \eqref{cond:IV_T3}, i.e., $(0,0)\notin \mathcal{T}_3$, $(0,1)\notin \mathcal{T}_3$, $(1,0)\notin \mathcal{T}_3$ and $\{(1,1),(2,0),(2,1)\}\bigcap $ $\mathcal{T}_3\neq \emptyset$. So we have
	$h^{\text{IV}}_3={K_1K_2-3 \choose t}-{K_1K_2-6\choose t}={9-3\choose 2}-{9-6\choose 2}=15-3=12$.
	Let us first consider $\mathcal{T}_3=\{(0,2),(1,1)\}$. The packets in coded signal
	$$
	X_{\{(1,0)\}\bigcup \mathcal{T}_3}=X_{\{(0,2),(1,0),(1,1)\}}=W_{d_{\mathcal{U}^{\text{IV}}_{0,2}},\{(1,0),(1,1)\}}\bigoplus W_{d_{\mathcal{U}^{\text{IV}}_{1,0}},\{(0,2),(1,1)\}}\bigoplus W_{d_{\mathcal{U}^{\text{IV}}_{1,1}},\{(0,2),(1,0)\}}
	$$
	can be retrieved by user $\mathcal{U}^{\text{IV}}_{1,0}$ through its connected cache nodes $\text{C}_{1,0}$ and $\text{C}_{1,1}$, then the server only needs to send the following modified coded signal
	$$
	X'_{\{(0,2),(1,0),(1,1)\}}=X_{\{(0,2),(1,0),(1,1)\}}-W_{d_{\mathcal{U}^{\text{IV}}_{1,0}},\{(0,2),(1,1)\}}=W_{d_{\mathcal{U}^{\text{IV}}_{0,2}},\{(1,0),(1,1)\}}\oplus W_{d_{\mathcal{U}^{\text{IV}}_{1,1}},\{(0,2),(1,0)\}}.
	$$
	From \eqref{eq-packets-IV}, user $\mathcal{U}^{\text{IV}}_{0,0}$ can retrieve all the packets in $X'_{\{(0,2),(1,0),(1,1)\}}$ from its connected cache node $\text{C}_{1,0}$. Similarly we can check that user $\mathcal{U}^{\text{IV}}_{0,0}$ can retrieve all the packets in the following modified coded signals
	\begin{align*}
		&X'_{\{(0,2),(1,0),(1,1)\}},\ X'_{\{(0,2),(1,0),(2,0)\}},\ X'_{\{(0,2),(1,0),(2,1)\}},\ X'_{\{(1,0),(1,1),(1,2)\}},\\
		& X'_{\{(1,0),(1,1),(2,0)\}},X'_{\{(1,0),(1,1),(2,1)\}},\ X'_{\{(1,0),(1,1),(2,2)\}},\ X'_{\{(1,0),(1,2),(2,0)\}},\\
		&X'_{\{(1,0),(1,2),(2,1)\}},\ X'_{\{(1,0),(2,0),(2,1)\}},\ X'_{\{(1,0),(2,0),(2,2)\}},\ X'_{\{(1,0),(2,1),(2,2)\}},
	\end{align*}
	which are generated by the following coded signals respectively
	\begin{align}
		&X_{\{(0,2),(1,0),(1,1)\}},\ X_{\{(0,2),(1,0),(2,0)\}},\ X_{\{(0,2),(1,0),(2,1)\}},\ X_{\{(1,0),(1,1),(1,2)\}},\nonumber\\
		&X_{\{(1,0),(1,1),(2,0)\}},\ X_{\{(1,0),(1,1),(2,1)\}},\ X_{\{(1,0),(1,1),(2,2)\}},\ X_{\{(1,0),(1,2),(2,0)\}},\nonumber\\
		&X_{\{(1,0),(1,2),(2,1)\}},\ X_{\{(1,0),(2,0),(2,1)\}},\ X_{\{(1,0),(2,0),(2,2)\}},\ X_{\{(1,0),(2,1),(2,2)\}}.\label{eq:IV-cancelsignal-S3}
	\end{align}
	
	\item Finally, we count the value of $h^{\text{IV}}_4$ and find the $h^{\text{IV}}_4$ coded signals. We can obtain the following $2$-subsets $\mathcal{T}_4$,
	\begin{align*}
		&\{(0,2),(1,2)\},\ \{(0,2),(2,1)\},\ \{(0,2),(2,2)\},\ \{(1,2),(2,0)\},\ \{(1,2),(2,1)\},\\
		&\{(1,2),(2,2)\},\ \{(2,0),(2,1)\},\ \{(2,0),(2,2)\},\ \{(2,1),(2,2)\},
	\end{align*}
	satisfying condition \eqref{cond:IV_T4}, i.e., $(0,0)\notin \mathcal{T}_4$, $(0,1)\notin \mathcal{T}_4$, $(1,0)\notin \mathcal{T}_4$, $(1,1)\notin \mathcal{T}_4$ and $\{(1,2),(2,1),(2,2)\}\bigcap \mathcal{T}_4\neq \emptyset$. So we have
	$h^{\text{IV}}_4={K_1K_2-4 \choose t}-{K_1K_2-7\choose t}={9-4\choose 2}-{9-7\choose 2}=10-1=9$.
	Let us first consider $\mathcal{T}_4=\{(0,2),(1,2)\}$. The packets in coded signal
	$$
	X_{\{(1,1)\}\bigcup \mathcal{T}_4}=X_{\{(0,2),(1,1),(1,2)\}}=W_{d_{\mathcal{U}^{\text{IV}}_{0,2}},\{(1,1),(1,2)\}}\bigoplus W_{d_{\mathcal{U}^{\text{IV}}_{1,1}},\{(0,2),(1,2)\}}\bigoplus W_{d_{\mathcal{U}^{\text{IV}}_{1,2}},\{(0,2),(1,1)\}}
	$$
	can be retrieved by user $\mathcal{U}^{\text{IV}}_{1,1}$ through its connected cache nodes $\text{C}_{1,1}$ and $\text{C}_{1,2}$, then the server only needs to send the following modified coded signal
	$$
	X'_{\{(0,2),(1,1),(1,2)\}}=X_{\{(0,2),(1,1),(1,2)\}}-W_{d_{\mathcal{U}^{\text{IV}}_{1,1}},\{(0,2),(1,2)\}}=W_{d_{\mathcal{U}^{\text{IV}}_{0,2}},\{(1,1),(1,2)\}}\bigoplus W_{d_{\mathcal{U}^{\text{IV}}_{1,2}},\{(0,2),(1,1)\}}.
	$$
	From \eqref{eq-packets-IV}, user $\mathcal{U}^{\text{IV}}_{0,0}$ can retrieve all the packets in $X'_{\{(0,2),(1,1),(1,2)\}}$ from the its connected cache node $\text{C}_{1,1}$. Similarly we can check that user $\mathcal{U}^{\text{IV}}_{0,0}$ can retrieve all the packets in modified coded signals
	\begin{align*}
		&X'_{\{(0,2),(1,1),(1,2)\}},\ X'_{\{(0,2),(1,1),(2,1)\}},\ X'_{\{(0,2),(1,1),(2,2)\}},\ X'_{\{(1,1),(1,2),(2,0)\}},\ X'_{\{(1,1),(1,2),(2,1)\}},\\
		&X'_{\{(1,1),(1,2),(2,2)\}},\ X'_{\{(1,1),(2,0),(2,1)\}},\ X'_{\{(1,1),(2,0),(2,2)\}},\ X'_{\{(1,1),(2,1),(2,2)\}},
	\end{align*}
	which are generated by the following coded signals respectively
	\begin{align}
		&X_{\{(0,2),(1,1),(1,2)\}},\ X_{\{(0,2),(1,1),(2,1)\}},\ X_{\{(0,2),(1,1),(2,2)\}},\ X_{\{(1,1),(1,2),(2,0)\}},\ X_{\{(1,1),(1,2),(2,1)\}},\nonumber\\
		&X_{\{(1,1),(1,2),(2,2)\}},\ X_{\{(1,1),(2,0),(2,1)\}},\ X_{\{(1,1),(2,0),(2,2)\}},\ X_{\{(1,1),(2,1),(2,2)\}}.\label{eq:IV-cancelsignal-S4}
	\end{align}
\end{itemize}

Obviously, the $h^{\text{IV}}_1$, $h^{\text{IV}}_2$, $h^{\text{IV}}_3$ and $h^{\text{IV}}_2$ coded signals in \eqref{eq:IV-cancelsignal-S1}, \eqref{eq:IV-cancelsignal-S2}, \eqref{eq:IV-cancelsignal-S3} and \eqref{eq:IV-cancelsignal-S4} are different from each other. So there are exactly $h^{\text{IV}}=h^{\text{IV}}_{1}+h^{\text{IV}}_{2}+h^{\text{IV}}_{3}+h^{\text{IV}}_{4}=54$ coded (or modified coded) signals which can be retrieved by user $\mathcal{U}^{\text{IV}}_{0,0}$.

Next, let us consider the transmission load of our two schemes when the value of $r$ is in different ranges.
\begin{itemize}
\item When $\frac{\sqrt{2}}{2} < r < 1$, there are $8$ sub-types of users in 2D caching-aided UDN system. From above description, we have $H=2h^{\text{II}}+4h^{\text{III}}+h^{\text{IV}}=2\times 13+4\times 33+54=212$. We use a $(84,71)$ MDS code, a $(84,51)$ MDS code and a $(84,30)$ MDS code for the $84$ coded signals sent to Type II, Type III and Type IV users respectively, the server only sends $8\times 84-H=672-212=460$ coded signals for all the users. Then the transmission load of our Scheme B is $R_{\text{B}}=\frac{460}{36}=12.78$ which is $0.68$ times smaller than the transmission load $R_{\text{A}}=\frac{8*84}{36}=18.67$ of Scheme A.

\item When $r=\frac{\sqrt{2}}{2}$, there are only Type I user and Type II user, total of $3$ sub-types of users in 2D caching-aided UDN system. From the above description, we have $H=2h^{\text{II}}=2\times 13=26$. We use a $(84,71)$ MDS code for the $84$ coded signals sent to Type II user, the server only sends $3\times 84-H=252-26=226$ coded signals for all the users. Then the transmission load of our Scheme B is $R_{\text{B}}=\frac{226}{36}=6.28$ which is smaller than the transmission load $R_{\text{A}}=\frac{3*84}{36}=7$ of Scheme A.

\item When $r = 1$, there are Type II user, Type III user and Type IV user, in total $7$ sub-types of users in 2D caching-aided UDN system. From above description, we have $H=2h^{\text{II}}+4h^{\text{III}}+h^{\text{IV}}=2\times 13+4\times 33+54=212$. We use a $(84,71)$ MDS code, a $(84,51)$ MDS code and a $(84,30)$ MDS code for the $84$ coded signals sent to Type II, Type III and Type IV users respectively, the server only sends $7\times 84-H=588-212=376$ coded signals for all the users. Then the transmission load of our Scheme B is $R_{\text{B}}=\frac{460}{36}=10.44$ which is $0.64$ times smaller than the transmission load $R_{\text{A}}=\frac{7*84}{36}=16.33$ of Scheme A.
\end{itemize}

\section{Conclusion}
\label{sec-conlusion}
In this paper, we investigated the 2D caching-aided UDN system. We first showed that all the possible users can be divided into four classes. Based on this classification, Scheme A is obtained by means of the placement strategy and delivery strategy of the MN scheme. By theoretical analysis, we showed that Scheme A is order optimal. To further delete the redundancy distributed in the transmitted signals of Scheme A, Scheme B which has a smaller transmission load was obtained. Finally, the numerical analysis shows that our schemes have significant advantages over the conventional uncoded scheme for the caching-aided UDN system.

\begin{appendices}
\section{The detailed proof of the user's types}
\label{appendix-user-type}
According to Cartesian coordinate system, we can easily check that the range of any cache node $\text{C}_{k_1,k_2}$ where $k_1\in [0:K_1),k_2\in [0:K_2)$ can be divided into the following four symmetric ranges.
\begin{align}
\label{eq-range-a-cycle}
\begin{split}
\{(k_1+x,k_2+x)\ |\ \sqrt{x^2+y^2}\leq r,x>0,y>0\}, \\
\{(k_1+x,k_2+x)\ |\ \sqrt{x^2+y^2}\leq r,x>0,y<0\}, \\
\{(k_1+x,k_2+x)\ |\ \sqrt{x^2+y^2}\leq r,x<0,y>0\}, \\
\{(k_1+x,k_2+x)\ |\ \sqrt{x^2+y^2}\leq r,x<0,y<0\}.
\end{split}
\end{align}So it is sufficient to consider one of the above four ranges for the whole service range of a cache node. Now let us consider the first range which is composed of the cache nodes $\text{C}_{k_1,k_2}$, $\text{C}_{k_1,k_2+1}$, $\text{C}_{k_1+1,k_2}$ and $\text{C}_{k_1+1,k_2+1}$. From \eqref{eq-righ-up-down-all}, we only need
\begin{align}
\label{eq-righ-up-down-part1}
\begin{split}
&e_{0,0}(x,y)=d_{(k_1+x,k_2+y),(k_1,k_2)}\ \ \ \ \ \ \ =\sqrt{x^2+y^2},\\
&e_{0,1}(x,y)=d_{(k_1+x,k_2+y),(k_1,k_2+1)}\ \ \ \ =\sqrt{x^2+(1-y)^2},\\
&e_{1,0}(x,y)=d_{(k_1+x,k_2+y),(k_1+1,k_2)}\ \ \ \ =\sqrt{(1-x)^2+y^2},\\
&e_{1,1}(x,y)=d_{(k_1+x,k_2+y),(k_1+1,k_2+1)}\ =\sqrt{(1-x)^2+(1-y)^2}.
\end{split}
\end{align}
According to the values of $e_{0,0}(x,y)$, $e_{0,1}(x,y)$, $e_{1,0}(x,y)$, $e_{1,1}(x,y)$ in \eqref{eq-righ-up-down-part1} and $\frac{\sqrt{2}}{2}< r < 1$, the users can be divided into the following four types.
\begin{itemize}
\item Type I: There exist some pairs $(x,y)\in \mathbb{R}^2$ satisfying all of the equations
\begin{eqnarray}
\label{eq-tyep-I-codidtion-1}
\begin{split}	
e_{0,0}(x,y)&=\sqrt{x^2+y^2}<r,\ \ \ \ \ \ \ \ \ e_{0,1}(x,y)=\sqrt{x^2+(1-y)^2}>r, \\
e_{1,0}(x,y)&=\sqrt{(1-x)^2+y^2}>r,\ \ e_{1,1}(x,y)=\sqrt{(1-x)^2+(1-y)^2}>r.
\end{split}
\end{eqnarray}For instance $x=y=\frac{1-r}{2}$ is the solution of the above equation \eqref{eq-tyep-I-codidtion-1}. In this case, the user $\text{U}_{k_1+x,k_2+y}$ where the pair $(x,y)$ satisfies \eqref{eq-tyep-I-codidtion-1} can only access cache node $\text{C}_{k_1,k_2}$, the user accessing cache node $\text{C}_{k_1,k_2}$ is denoted by $\text{U}\{(k_1,k_2)\}$. Similarly, we can also denote all users in the four symmetric ranges in \eqref{eq-range-a-cycle}, and thus the Type I user can be represented in \eqref{eq-type-I}.
\item Type II: There exist some pairs $(x,y)\in \mathbb{R}^2$ satisfying all of the equations
\begin{eqnarray}
\label{eq-tyep-II-codidtion-1}
\begin{split}
e_{0,0}(x,y)&=\sqrt{x^2+y^2}<r,\ \ \ \ \ \ \ \ \ e_{0,1}(x,y)=\sqrt{x^2+(1-y)^2}<r, \\
e_{1,0}(x,y)&=\sqrt{(1-x)^2+y^2}>r,\ \ e_{1,1}(x,y)=\sqrt{(1-x)^2+(1-y)^2}>r.
\end{split}
\end{eqnarray}
For instance when $(x,y)=(0,\frac{1}{2})$, the above equations in \eqref{eq-tyep-II-codidtion-1} hold. The user $\text{U}_{k_1+x,k_2+y}$ where the pair $(x,y)$ satisfies \eqref{eq-tyep-II-codidtion-1} can exactly access the cache nodes $\text{C}_{k_1,k_2}$ and $\text{C}_{k_1,<k_2+1>_{K_2}}$, we denote the user accessing cache nodes $\text{C}_{k_1,k_2}$ and $\text{C}_{k_1,<k_2+1>_{K_2}}$ by $\text{U}\{(k_1,k_2),(k_1,<k_2+1>_{K_2})\}$. Similarly, we can also denote all the users in the four symmetric ranges in \eqref{eq-range-a-cycle}, and thus the Type II-1 user accessing two consecutive cache nodes $\text{C}_{k_1,k_2}$ and $\text{C}_{k_1,<k_2+1>_{K_2}}$ in the horizontal direction can be summarized as \eqref{eq-type-II-1}.

Similar to \eqref{eq-tyep-II-codidtion-1}, there also exist some pairs $(x,y)\in\mathbb{R}^2$ satisfying all of the equations
\begin{eqnarray}
	\label{eq-tyep-II-codidtion-2}
	\begin{split}
	e_{0,0}(x,y)&=\sqrt{x^2+y^2}<r,\ \ \ \ \ \ \ \ \ e_{0,1}(x,y)=\sqrt{x^2+(1-y)^2}>r, \\
	e_{1,0}(x,y)&=\sqrt{(1-x)^2+y^2}<r,\ \ e_{1,1}(x,y)=\sqrt{(1-x)^2+(1-y)^2}>r.
	\end{split}
\end{eqnarray} For instance when $(x,y)=(\frac{1}{2},0)$, the above equations \eqref{eq-tyep-II-codidtion-2} hold respectively. Similar to the Type II-1 user introduction, the Type II-2 user accessing two consecutive cache nodes $\text{C}_{k_1,k_2}$ and $\text{C}_{<k_1 + 1>_{K_1},k_2}$ in vertical direction are denoted by \eqref{eq-type-II-2}.

\item Type III: There exists some pairs $(x,y) \in \mathbb{R}^2$ satisfying all of the equations
\begin{eqnarray}
	\label{eq-tyep-III-codition-1}
	\begin{split}
	e_{0,0}(x,y)&=\sqrt{x^2+y^2}<r,\ \ \ \ \ \ \ \ \ e_{0,1}(x,y)=\sqrt{x^2+(1-y)^2}<r, \\
	e_{1,0}(x,y)&=\sqrt{(1-x)^2+y^2}<r,\ \ e_{1,1}(x,y)=\sqrt{(1-x)^2+(1-y)^2}>r.
	\end{split}
\end{eqnarray}
For instance when $(x,y)=(1-\sqrt{r^2-\frac{1}{4}}, 1-\sqrt{r^2-\frac{1}{4}})$, the above equations in \eqref{eq-tyep-III-codition-1} hold. The user $\text{U}_{k_1+x,k_2+y}$ where the pair $(x,y)$ satisfies \eqref{eq-tyep-III-codition-1} can exactly access the cache nodes $\text{C}_{k_1,k_2}$, $\text{C}_{<k_1+1>_{K_1},k_2}$ and $\text{C}_{k_1,<k_2+1>_{K_2}}$, we denote the user accessing $\text{C}_{k_1,k_2}$, $\text{C}_{<k_1+1>_{K_1},k_2}$ and $\text{C}_{k_1,<k_2+1>_{K_2}}$ by $\text{U}\{(k_1,k_2),(<k_1+1>_{K_1},k_2),(k_1,<k_2+1>_{K_2})\}$. Similarly, we can also denote all the users in the four symmetric ranges in \eqref{eq-range-a-cycle}, and thus the Type III-1 user accessing three cache nodes $\text{C}_{k_1,k_2}$, $\text{C}_{<k_1+1>_{K_1},k_2}$ and $\text{C}_{k_1,<k_2+1>_{K_2}}$ are represented as \eqref{eq-type-III-1}.

Similar to \eqref{eq-tyep-III-codition-1}, there also exists some pairs $(x,y) \in \mathbb{R}^2$ satisfying all of the equations
\begin{eqnarray}
	\label{eq-tyep-III-codition-2}
	\begin{split}
	e_{0,0}(x,y)&=\sqrt{x^2+y^2}<r,\ \ \ \ \ \ \ \ \ e_{0,1}(x,y)=\sqrt{x^2+(1-y)^2}>r, \\
	e_{1,0}(x,y)&=\sqrt{(1-x)^2+y^2}<r,\ \ e_{1,1}(x,y)=\sqrt{(1-x)^2+(1-y)^2}<r.
	\end{split}
\end{eqnarray}
For instance, when $(x,y)=(\sqrt{r^2-\frac{1}{4}}, 1-\sqrt{r^2-\frac{1}{4}})$, the above equations in \eqref{eq-tyep-III-codition-2} hold. Similar to the Type III-1 user introduction, the Type III-2 user accessing three cache nodes $\text{C}_{k_1,k_2}$, $\text{C}_{<k_1+1>_{K_1},k_2}$ and $\text{C}_{<k_1+1>_{K_1},<k_2+1>_{K_2}}$ are represented as \eqref{eq-type-III-2}.

Similar to \eqref{eq-tyep-III-codition-1} and \eqref{eq-tyep-III-codition-2}, there also exists some pairs $(x,y) \in \mathbb{R}^2$ satisfying all of the equations
\begin{eqnarray}
	\label{eq-tyep-III-codition-3}
	\begin{split}
	e_{0,0}(x,y)&=\sqrt{x^2+y^2}<r,\ \ \ \ \ \ \ \ \ e_{0,1}(x,y)=\sqrt{x^2+(1-y)^2}<r, \\
	e_{1,0}(x,y)&=\sqrt{(1-x)^2+y^2}>r,\ \ e_{1,1}(x,y)=\sqrt{(1-x)^2+(1-y)^2}<r.
	\end{split}
\end{eqnarray}
For instance, when $(x,y)=(1-\sqrt{r^2-\frac{1}{4}}, \sqrt{r^2-\frac{1}{4}})$, the above equations in \eqref{eq-tyep-III-codition-3} hold. Similar to the above introduction, the Type III-3 user accessing three cache nodes $\text{C}_{k_1,k_2}$, $\text{C}_{k_1,<k_2+1>_{K_2}}$ and $\text{C}_{<k_1+1>_{K_1},<k_2+1>_{K_2}}$ are represented as \eqref{eq-type-III-3}.

Since the type of user in the first range is symmetrical with that in the fourth range from \eqref{eq-range-a-cycle}, the Type III-4 user in the fourth range are also symmetrical with the Type III-1 user in the first range, then the Type III-4 user located at cache node $\text{C}_{k_1,k_2}$ serving rang can be represented in the fourth range. In the fourth range which is composed of the cache nodes $\text{C}_{k_1,k_2}$, $\text{C}_{k_1,<k_2-1>_{K_2}}$, $\text{C}_{<k_1-1>_{K_1},k_2}$ and $\text{C}_{<k_1-1>_{K_1},<k_2-1>_{K_2}}$ where $k_1\in [0:K_1), k_2\in [0:K_2)$, according to the values of $e_{0,0}(x,y)$, $e_{0,-1}(x,y)$, $e_{-1,0}(x,y)$ and $e_{-1,-1}(x,y)$ in \eqref{eq-righ-up-down-all}, there exists some pairs $(x,y) \in \mathbb{R}^2$ satisfying all of the equations
\begin{eqnarray}
	\label{eq-tyep-III-codition-4}
	\begin{split}
	e_{0,0}(x,y)&=\sqrt{x^2+y^2}<r,\ \ \ \ \ \ \ \ \ e_{0,-1}(x,y)=\sqrt{x^2+(1+y)^2}<r, \\
	e_{-1,0}(x,y)&=\sqrt{(1+x)^2+y^2}<r,\ \ e_{-1,-1}(x,y)=\sqrt{(1+x)^2+(1+y)^2}>r.
	\end{split}
\end{eqnarray}
For instance, when $(x,y)=(-1+\sqrt{r^2-\frac{1}{4}}, -1+\sqrt{r^2-\frac{1}{4}})$, the above equations in \eqref{eq-tyep-III-codition-4} hold.The user $\text{U}_{k_1+x,k_2+y}$ where the pair $(x,y)$ satisfies \eqref{eq-tyep-III-codition-4} can exactly access the cache nodes $\text{C}_{k_1,k_2}$, $\text{C}_{k_1,<k_2-1>_{K_2}}$ and $\text{C}_{<k_1-1>_{K_1},k_2}$, we denote the user accessing $\text{C}_{k_1,k_2}$, $\text{C}_{k_1,<k_2-1>_{K_2}}$ and $\text{C}_{<k_1-1>_{K_1},k_2}$ by $\text{U}\{(k_1,k_2),(k_1,<k_2-1>_{K_2}),(<k_1-1>_{K_1},k_2)\}$. Similarly, we can also denote all the users in the four symmetric ranges in \eqref{eq-range-a-cycle}, and thus type III-4 user accessing three cache nodes $\text{C}_{k_1,k_2}$, $\text{C}_{k_1,<k_2-1>_{K_2}}$ and $\text{C}_{<k_1-1>_{K_1},k_2}$ are represented as \eqref{eq-type-III-4}.

\item Type IV: There exists some pairs $(x,y) \in \mathbb{R}^2$ satisfying all of the equations
\begin{eqnarray}
	\label{eq-tyep-IV-codition-1}
	\begin{split}
	e_{0,0}(x,y)&=\sqrt{x^2+y^2}<r,\ \ \ \ \ \ \ \ \ e_{0,1}(x,y)=\sqrt{x^2+(1- y)^2}<r, \\
	e_{1,0}(x,y)&=\sqrt{(1- x)^2+y^2}<r,\ \ e_{1,1}(x,y)=\sqrt{(1- x)^2+(1- y)^2}<r.
	\end{split}
\end{eqnarray}
For instance, when $x=y=\frac{1}{2}$, the above equations in \eqref{eq-tyep-IV-codition-1} hold. In this case, the user $\text{U}_{k_1+x,k_2+y}$ where the pair $(x,y)$ satisfies \eqref{eq-tyep-IV-codition-1} can exactly access four cache nodes $\text{C}_{k_1,k_2}$, $\text{C}_{<k_1+1>_{K_1},k_2}$, $\text{C}_{k_1,<k_2+1>_{K_2}}$ and $\text{C}_{<k_1+1>_{K_1},<k_2+1>_{K_2}}$, and is denoted by $\text{U}(k_1,k_2),(<k_1+1>_{K_1},k_2),(k_1,<k_2+1>_{K_2}),(<k_1+1>_{K_1},<k_2+1>_{K_2})$. Similarly, we can also denote all the users in the four symmetric ranges in \eqref{eq-range-a-cycle}, and thus the Type IV user can be represented as \eqref{eq-type-IV}.
\end{itemize}

\section{The proof of the coded signals and modified coded signals for Type III user}
\label{appendix-Type III_h}
Let us consider the users in Type III-1 first. For any integer pair $(k_1,k_2)\in \mathcal{K}$, the following statements can be obtained by the accessing topology.
\begin{itemize}
\item User $\mathcal{U}^{\text{III-1}}_{k_1,k_2}=$ $\{(k_1,k_2),(k_1,<k_2+1>_{K_2}),(<k_1+1>_{K_1},k_2)\}$ can access cache nodes $\text{C}_{k_1,k_2}$, $\text{C}_{k_1,<k_2+1>_{K_2}}$ and $\text{C}_{<k_1+1>_{K_1},k_2}$;
\item User $\mathcal{U}^{\text{III-1}}_{k_1,<k_2+1>_{K_2}}=$ $\{(k_1,<k_2+1>_{K_2}),(k_1,<k_2+2>_{K_2}),(<k_1+1>_{K_1},<k_2+1>_{K_2})\}$ can access cache nodes $\text{C}_{k_1,<k_2+1>_{K_2}}$, $\text{C}_{k_1,<k_2+2>_{K_2}}$ and $\text{C}_{<k_1+1>_{K_1},<k_2+1>_{K_2}}$;
\item User $\mathcal{U}^{\text{III-1}}_{<k_1+1>_{K_1},k_2}=$ $\{(<k_1+1>_{K_1},k_2),(<k_1+1>_{K_1},<k_2+1>_{K_2}),(<k_1+2>_{K_1},k_2)\}$ can access cache nodes $\text{C}_{<k_1+1>_{K_1},k_2}$, $\text{C}_{<k_1+1>_{K_1},<k_2+1>_{K_2}}$ and $\text{C}_{<k_1+2>_{K_1},k_2}$.
\end{itemize}
Let us see the delivery strategy for the users $\mathcal{U}^{\text{III-1}}_{k_1,k_2}$, $\mathcal{U}^{\text{III-1}}_{k_1,<k_2+1>_{K_2}}$ and $\mathcal{U}^{\text{III-1}}_{<k_1+1>_{K_1},k_2}$ of Scheme A by Line 11 of Algorithm \ref{alg-MN}.

\begin{itemize}
\item Let us consider the value of $h^{\text{III}}_1$ by counting the number of $t$-subset $\mathcal{T}_1\in {\mathcal{K} \choose t}$ satisfying the conditions
\begin{align}
\label{cond:III-1_T1}
&\ \ \ \ (k_1,k_2)\notin \mathcal{T}_1,\ \ \ \ \text{and }\ \ \ (\mathcal{U}^{\text{III-1}}_{k_1,k_2}\setminus\{(k_1,k_2)\})\nonumber\\
&=\{(k_1,<k_2+1>_{K_2}),(<k_1+1>_{K_1},k_2)\}\bigcap \mathcal{T}_1 \neq \emptyset.
\end{align}
Let $\mathcal{S}_1=\{(k_1,k_2)\}\bigcup \mathcal{T}_1$. From \eqref{eq-type-III-1} and \eqref{SC1-Z_Uk}, we have $(\mathcal{S}_1\setminus\{(k_1',k_2')\})\bigcap \mathcal{U}^{\text{III-1}}_{k_1,k_2}\neq \emptyset$ where $(k_1',k_2')\in \mathcal{S}_1 $. Then the transmitted coded signal
$X_{\mathcal{S}_1}=\bigoplus_{(k_1',k_2')\in \mathcal{S}_1}W_{d_{\mathcal{U}_{k_1',k_2'}},\mathcal{S}_1\setminus\{(k_1',k_2')\}}$
can be retrieved by user $\mathcal{U}^{\text{III-1}}_{k_1,k_2}$, i.e., the packets in $X_{\mathcal{S}_1}$ can be retrieved by user $\mathcal{U}^{\text{III-1}}_{k_1,k_2}$ through cache nodes $\text{C}_{k_1,k_2}$, $\text{C}_{k_1,<k_2+1>_{K_2}}$ and $\text{C}_{<k_1+1>_{K_1},k_2}$. We can check that there are
\begin{align}
\label{eq-redundancy-III-1_h1}
h^{\text{III}}_{1}={K_1K_2-1 \choose t}-{K_1K_2-3 \choose t}
\end{align}
such subsets $\mathcal{T}_1$. So there are $h^{\text{III}}_{1}$ such transmitted coded signals in Line 11 of Algorithm \ref{alg-MN} which can be retrieved by user $\mathcal{U}^{\text{III-1}}_{k_1,k_2}$.
	
\item Let us consider the value of $h^{\text{III}}_{2}$ by counting the number of $t$-subset $\mathcal{T}_2\in {\mathcal{K} \choose t}$ satisfying the conditions
\begin{align}
\label{cond:III-1_T2}	
&(k_1,k_2) \notin \mathcal{T}_2,\ \ \ \ \ (k_1,<k_2+1>_{K_2}) \notin \mathcal{T}_2\ \ \ \ \ \ \ \text{and } \nonumber \\ &(\mathcal{U}^{\text{III-1}}_{k_1,<k_2+1>_{K_2}}\setminus \{(k_1,<k_2+1>_{K_2})\})
=\{(k_1,<k_2+2>_{K_2}),\nonumber \\
&(<k_1+1>_{K_1},<k_2+1>_{K_2})\} \bigcap \mathcal{T}_2\neq \emptyset.
\end{align}
Let $\mathcal{S}_2=\{(k_1,<k_2+1>_{K_2})\}\bigcup \mathcal{T}_2$. From \eqref{eq-type-III-1} and \eqref{SC1-Z_Uk}, we have $(\mathcal{S}_2\setminus\{(k_1',k_2')\})\bigcap$ $\mathcal{U}^{\text{III-1}}_{k_1,<k_2+1>_{K_2}} \neq \emptyset$ where $(k_1',k_2')\in \mathcal{S}_2 $, then the transmitted coded signal
$X_{\mathcal{S}_2}=\bigoplus_{(k_1',k_2')\in \mathcal{S}_2}$ $W_{d^{\text{III-1}}_{\mathcal{U}_{k_1',k_2'}},\mathcal{S}_2\setminus\{(k_1',k_2')\}}$
can be retrieved by user $\mathcal{U}^{\text{III-1}}_{k_1,<k_2+1>_{K_2}}$, i.e., the packets in $X_{\mathcal{S}_2}$ can be retrieved by user $\mathcal{U}^{\text{III-1}}_{k_1,<k_2+1>_{K_2}}$ through cache nodes $\text{C}_{k_1,<k_2+1>_{K_2}}$, $\text{C}_{k_1,<k_2+2>_{K_2}}$ and $\text{C}_{<k_1+1>_{K_1},<k_2+1>_{K_2}}$. Then the server only needs to send the following modified coded signal
$X'_{\mathcal{S}_2}=X_{\mathcal{S}_2}-W_{d_{\mathcal{U}^{\text{III-1}}_{k_1,<k_2+1>_{K_2}}},\mathcal{T}_2}=\bigoplus_{(k'_1,k'_2)\in (\mathcal{S}_2\setminus(k_1,<k_2+1>_{K_2}))}$ $W_{d_{\mathcal{U}^{\text{III-1}}_{k'_1,k'_2}},\mathcal{S}_2\setminus\{(k'_1,k'_2)\}}$.
We can check that user $\mathcal{U}^{\text{III-1}}_{k_1,k_2}$ can retrieve all the packets in $X'_{\mathcal{S}_2}$ since it can retrieve all the packets labeled by $\mathcal{S}_2\setminus\{(k'_1,k'_2)\}$ where $(k'_1,k'_2)\neq (k_1,<k_2+1>_{K_2})$ from \eqref{SC1-Z_Uk}. We can check that there are
\begin{align}
\label{eq-redundancy-III-1_h2}
h^{\text{III}}_{2}={K_1K_2-2 \choose t}-{K_1K_2-4 \choose t}
\end{align}
such subsets $\mathcal{T}_2$. So there are $h^{\text{III}}_{2}$ such transmitted modified coded signals in Line 11 of Algorithm \ref{alg-MN} which can be retrieved by user $\mathcal{U}^{\text{III-1}}_{k_1,k_2}$.
	
\item Let us consider the value of $h^{\text{III}}_{3}$ by counting the number of $t$-subset $\mathcal{T}_3\in {\mathcal{K} \choose t}$ satisfying the conditions
\begin{align}
\label{cond:III-1_T3}
&(k_1,k_2) \notin \mathcal{T}_3,\ (k_1,<k_2+1>_{K_2}) \notin \mathcal{T}_3,\ (<k_1+1>_{K_1},k_2) \notin \mathcal{T}_3\ \text{and } \nonumber \\
&\ \ \ \ (\mathcal{U}^{\text{III-1}}_{<k_1+1>_{K_1},k_2}\setminus \{(<k_1+1>_{K_1},k_2)\})\ =\ \{(<k_1+2>_{K_1},k_2),\nonumber \\
&\ \ \  \ \ \ \ \ \ \ \ \ \ (<k_1+1>_{K_1},<k_2+1>_{K_2})\} \bigcap \mathcal{T}_3\neq \emptyset.\
\end{align}
Let $\mathcal{S}_3=\{(<k_1+1>_{K_1},k_2)\}\bigcup \mathcal{T}_3$. From \eqref{eq-type-III-1} and \eqref{SC1-Z_Uk}, we have $(\mathcal{S}_3\setminus\{(k_1',k_2')\})\bigcap$ $\mathcal{U}^{\text{III-1}}_{<k_1+1>_{K_1},k_2} \neq \emptyset$ where $(k_1',k_2')\in \mathcal{S}_3 $, then transmitted coded signal
$X_{\mathcal{S}_3}=\bigoplus_{(k_1',k_2')\in \mathcal{S}_3}$ $W_{d^{\text{III-1}}_{\mathcal{U}_{k_1',k_2'}},\mathcal{S}_3\setminus\{(k_1',k_2')\}}$
can be retrieved by user $\mathcal{U}^{\text{III-1}}_{<k_1+1>_{K_1},k_2}$, i.e., the packers in $X_{\mathcal{S}_3}$ can be retrieved by user $\mathcal{U}^{\text{III-1}}_{<k_1+1>_{K_1},k_2}$ through cache nodes $\text{C}_{<k_1+1>_{K_1},k_2}$, $\text{C}_{<k_1+1>_{K_1},<k_2+1>_{K_2}}$ and $\text{C}_{<k_1+2>_{K_1},k_2}$. Then the server only needs to send the following modified coded signal
$X'_{\mathcal{S}_3}=X_{\mathcal{S}_3}-W_{d_{\mathcal{U}^{\text{III-1}}_{<k_1+1>_{K_1},k_2}},\mathcal{T}_3}=\bigoplus_{(k'_1,k'_2)\in (\mathcal{S}_3\setminus\{(<k_1+1>_{K_1},k_2)\})}$ $W_{d_{\mathcal{U}^{\text{III-1}}_{k'_1,k'_2}},\mathcal{S}_3\setminus\{(k'_1,k'_2)\}}$.
We can check that user $\mathcal{U}^{\text{III-1}}_{k_1,k_2}$ can retrieve all the packets in $X'_{\mathcal{S}_3}$ since it can retrieve all the packets labeled by $\mathcal{S}_3\setminus\{(k'_1,k'_2)\}$ where $(k'_1,k'_2)\neq (<k_1+1>_{K_1},k_2)$ from \eqref{SC1-Z_Uk}. We can check that there are
\begin{align}
\label{eq-redundancy-III-1_h3}
h^{\text{III}}_{3}={K_1K_2-3 \choose t}-{K_1K_2-5 \choose t}
\end{align}
such subsets $\mathcal{T}_3$. So there are $h^{\text{III}}_{3}$ such transmitted modified coded signals in Line 11 of Algorithm \ref{alg-MN} which can be retrieved by user $\mathcal{U}^{\text{III-1}}_{k_1,k_2}$.
\end{itemize}	
Finally, by the assumptions $(k_1,k_2)\in \mathcal{S}_1$, $(k_1,k_2)\notin \mathcal{S}_2$, $(k_1,k_2)\notin \mathcal{S}_3$, $(k_1,<k_2+1>_{K_2})\in \mathcal{S}_2$, and $(k_1,<k_2+1>_{K_2})\notin \mathcal{S}_3$, we can verify $\mathcal{S}_1\neq \mathcal{S}_2$, $\mathcal{S}_1\neq \mathcal{S}_3$ and $\mathcal{S}_2\neq \mathcal{S}_3$. So there are exactly $h^{\text{III}}=h^{\text{III}}_{1}+h^{\text{III}}_{2}+h^{\text{III}}_{3}$ coded (or modified coded) signals which can be retrieved by user $\mathcal{U}^{\text{III-1}}_{k_1,k_2}$.

Similar to the discussion of Type III-1, for the users in Type III-2, Type III-3, Type III-4, we can also derive the values of $h^{\text{III}}_{1}$ in \eqref{eq-redundancy-III-1_h1}, $h^{\text{III}}_{2}$ in \eqref{eq-redundancy-III-1_h2} and $h^{\text{III}}_{3}$ in \eqref{eq-redundancy-III-1_h3} by discussing the following three classes of users
\begin{align*}
&\left\{\mathcal{U}^{\text{III-2}}_{k_1,k_2},\ \mathcal{U}^{\text{III-2}}_{<k_1+1>_{K_1},k_2},\ \mathcal{U}^{\text{III-2}}_{<k_1+1>_{K_1},<k_2+1>_{K_2}}\right\},\\
&\left\{\mathcal{U}^{\text{III-3}}_{k_1,k_2},\ \mathcal{U}^{\text{III-3}}_{k_1,<k_2+1>_{K_2}},\ \mathcal{U}^{\text{III-3}}_{<k_1+1>_{K_1},<k_2+1>_{K_2}}\right\},\\
&\left\{\mathcal{U}^{\text{III-4}}_{k_1,k_2},\ \mathcal{U}^{\text{III-4}}_{<k_1-1>_{K_1},k_2},\ \mathcal{U}^{\text{III-4}}_{<k_1-1>_{K_1},<k_2-1>_{K_2}}\right\}
\end{align*} respectively.

\section{The proof of the coded signals and modified coded signals for Type IV user}
\label{appendix-Type IV_h}
Let us consider the users in Type IV. For any integer pair $(k_1,k_2)\in \mathcal{K}$, the following statements can be obtained by the accessing topology.
\begin{itemize}
\item User $\mathcal{U}^{\text{IV}}_{k_1,k_2}=$ $\{(k_1,k_2),(k_1,<k_2+1>_{K_2})$, $(<k_1+1>_{K_1},k_2)$, $(<k_1+1>_{K_1},<k_2+\\1>_{K_2})\}$ can access cache nodes $\text{C}_{k_1,k_2}$, $\text{C}_{k_1,<k_2+1>_{K_2}}$, $\text{C}_{<k_1+1>_{K_1},k_2}$ and $\text{C}_{<k_1+1>_{K_1},<k_2+1>_{K_2}}$;
\item User $\mathcal{U}^{\text{IV}}_{k_1,<k_2+1>_{K_2}}=$ $\{(k_1,<k_2+1>_{K_2})$, $(k_1,<k_2+2>_{K_2})$, $(<k_1+1>_{K_1},<k_2+1>_{K_2})$, $(<k_1+1>_{K_1}$, $<k_2+2>_{K_2})\}$ can access cache nodes $\text{C}_{k_1,<k_2+1>_{K_2}}$, $\text{C}_{k_1,<k_2+2>_{K_2}}$, $\text{C}_{<k_1+1>_{K_1},<k_2+1>_{K_2}}$ and $\text{C}_{<k_1+1>_{K_1},<k_2+2>_{K_2}}$;
\item User $\mathcal{U}^{\text{IV}}_{<k_1+1>_{K_1},k_2}=$ $\{(<k_1+1>_{K_1},k_2)$, $(<k_1+1>_{K_1},<k_2+1>_{K_2})$, $(<k_1+2>_{K_1},k_2)$, $(<k_1+2>_{K_1},<k_2+1>_{K_2})\}$ can access cache nodes $\text{C}_{<k_1+1>_{K_1},k_2}$, $\text{C}_{<k_1+1>_{K_1},<k_2+1>_{K_2}}$, $\text{C}_{<k_1+2>_{K_1},k_2}$ and $\text{C}_{<k_1+2>_{K_1},<k_2+1>_{K_2}}$;
\item User $\mathcal{U}^{\text{IV}}_{<k_1+1>_{K_1},<k_2+1>_{K_2}}=$ $\{(<k_1+1>_{K_1},<k_2+1>_{K_2})$, $(<k_1+1>_{K_1},<k_2+2>_{K_2})$, $(<k_1+2>_{K_1},<k_2+1>_{K_2})$, $(<k_1+2>_{K_1},<k_2+2>_{K_2})\}$ can access cache nodes $\text{C}_{<k_1+1>_{K_1},<k_2+1>_{K_2}}$, $\text{C}_{<k_1+1>_{K_1},<k_2+2>_{K_2}}$, $\text{C}_{<k_1+2>_{K_1},<k_2+1>_{K_2}}$ and $\text{C}_{<k_1+2>_{K_1},<k_2+2>_{K_2}}$.
\end{itemize}

Let us see the delivery strategy for the users $\mathcal{U}^{\text{IV}}_{k_1,k_2}$. $\mathcal{U}^{\text{IV}}_{k_1,<k_2+1>_{K_2}}$, $\mathcal{U}^{\text{IV}}_{<k_1+1>_{K_1},k_2}$ and\\ $\mathcal{U}^{\text{IV}}_{<k_1+1>_{K_1},<k_2+1>_{K_2}}$ of Scheme A by Line 11 of Algorithm \ref{alg-MN}.
\begin{itemize}
\item Firstly, we consider the value of $h^{\text{IV}}_{1}$ by counting the number of $t$-subset $\mathcal{T}_1\in {\mathcal{K} \choose t}$ satisfying the conditions
\begin{align}
\label{cond:IV_T1}
&(k_1,k_2)\notin \mathcal{T}_1, \ \text{and}\ \ (\mathcal{U}^{\text{IV}}_{k_1,k_2}\setminus\{(k_1,k_2)\})=\{(k_1,<k_2+1>_{K_2}),\nonumber\\
&(<k_1+1>_{K_1},k_2),(<k_1+1>_{K_1},<k_2+1>_{K_2})\}\bigcap \mathcal{T}_1 \neq \emptyset.
\end{align}
Let $\mathcal{S}_1=\{(k_1,k_2)\}\bigcup \mathcal{T}_1$. From \eqref{eq-type-IV} and \eqref{SC1-Z_Uk}, we have $(\mathcal{S}_1\setminus\{(k_1',k_2')\})\bigcap \mathcal{U}^{\text{IV}}_{k_1,k_2}\neq \emptyset$ where $(k_1',k_2')\in \mathcal{S}_1$, then the transmitted coded signal
$X_{\mathcal{S}_1}=\bigoplus_{(k_1',k_2')\in \mathcal{S}_1}$ $W_{d_{\mathcal{U}_{k_1',k_2'}},\mathcal{S}_1\setminus\{(k_1',k_2')\}}$
can be retrieved by user $\mathcal{U}^{\text{IV}}_{k_1,k_2}$, i.e., the packets decoded from $X_{\mathcal{S}_1}$ can be retrieved by user $\mathcal{U}^{\text{IV}}_{k_1,k_2}$ through cache nodes $\text{C}_{k_1,k_2}$, $\text{C}_{k_1,<k_2+1>_{K_2}}$, $\text{C}_{<k_1+1>_{K_1},k_2}$ and $\text{C}_{<k_1+1>_{K_1},<k_2+1>_{K_2}}$. We can check that there are
$h^{\text{IV}}_{1}={K_1K_2-1 \choose t}-{K_1K_2-4 \choose t}$
such subsets $\mathcal{T}_1$, so there are $h^{\text{IV}}_{1}$ such transmitted signals in Line 11 of Algorithm \ref{alg-MN} which can be retrieved by user $\mathcal{U}^{\text{IV}}_{k_1,k_2}$.
	
\item Now we discuss the value $h^{\text{IV}}_2$ by counting the number of $t$-subset $\mathcal{T}_2\in {\mathcal{K} \choose t}$ satisfying the conditions
\begin{align}
\label{cond:IV_T2}	
&(k_1,k_2) \notin \mathcal{T}_2, \ \ \ \ (k_1,<k_2+1>_{K_2}) \notin \mathcal{T}_2\ \ \ \ \text{and }\nonumber\\
&   (\mathcal{U}^{\text{IV}}_{k_1,<k_2+1>_{K_2}}\setminus\{(k_1,<k_2+1>_{K_2})\})=\{(k_1,<k_2+2>_{K_2}),\nonumber\\
&(<k_1+1>_{K_1},<k_2+1>_{K_2}),(<k_1+1>_{K_1},<k_2+2>_{K_2})\} \bigcap \mathcal{T}_2\neq \emptyset.
\end{align}	
Let $\mathcal{S}_2=\{(k_1,<k_2+1>_{K_2})\}\bigcup \mathcal{T}_2$. From \eqref{eq-type-IV} and \eqref{SC1-Z_Uk}, we have $(\mathcal{S}_2\setminus\{(k_1',k_2')\})\bigcap$ $\mathcal{U}^{\text{IV}}_{k_1,<k_2+1>_{K_2}} \neq \emptyset$ where $(k_1',k_2')\in \mathcal{S}_2$, then the transmitted coded signal
$X_{\mathcal{S}_2}=\bigoplus_{(k_1',k_2')\in \mathcal{S}_2}$ $W_{d^{\text{IV}}_{\mathcal{U}_{k_1',k_2'}},\mathcal{S}_2\setminus\{(k_1',k_2')\}}$
can be retrieved by user $\mathcal{U}^{\text{IV}}_{k_1,<k_2+1>_{K_2}}$, i.e., the packets contained in $X_{\mathcal{S}_2}$ can be retrieved by user $\mathcal{U}^{\text{IV}}_{k_1,<k_2+1>_{K_2}}$ through cache nodes $\text{C}_{k_1,<k_2+1>_{K_2}}$, $\text{C}_{k_1,<k_2+2>_{K_2}}$, $\text{C}_{<k_1+1>_{K_1},<k_2+1>_{K_2}}$ and $\text{C}_{<k_1+1>_{K_1},<k_2+2>_{K_2}}$. Then the server only needs to send the modified coded signal
$
X'_{\mathcal{S}_2}=X_{\mathcal{S}_2}-W_{d_{\mathcal{U}^{\text{IV}}_{k_1,<k_2+1>_{K_2}}},\mathcal{T}_2}=\bigoplus_{(k'_1,k'_2)\in (\mathcal{S}_2\setminus\{(k_1,<k_2+1>_{K_2})\})}$\\ $W_{d_{\mathcal{U}^{\text{IV}}_{k'_1,k'_2}},\mathcal{S}_2\setminus\{(k'_1,k'_2)\}}$. We can check that user $\mathcal{U}^{\text{IV}}_{k_1,k_2}$ can retrieve all the packets in $X'_{\mathcal{S}_2}$ since it can retrieve all the packets labeled by $(\mathcal{S}_2\setminus\{(k'_1,k'_2)\})$ where $(k'_1,k'_2)\neq (k_1,<k_2+1>_{K_2})$ from \eqref{SC1-Z_Uk}. We can check that there are
$h^{\text{IV}}_{2}={K_1K_2-2 \choose t}-{K_1K_2-5 \choose t}$
such subset $\mathcal{T}_2$, so	there are $h^{\text{IV}}_{2}$ such transmitted signals in Line 11 of Algorithm \ref{alg-MN} which can be retrieved by user $\mathcal{U}^{\text{IV}}_{k_1,k_2}$.
	
\item Next, we consider the value $h^{\text{IV}}_3$ by counting the number of $t$-subset $\mathcal{T}_3\in {\mathcal{K} \choose t}$ satisfying the conditions
\begin{align}
\label{cond:IV_T3}
&(k_1,k_2) \notin \mathcal{T}_3,\ \  (k_1,<k_2+1>_{K_2}) \notin \mathcal{T}_3,\ \  (<k_1+1>_{K_1},k_2) \notin \mathcal{T}_3,\ \  \text{and}\nonumber \\
&(\mathcal{U}^{\text{IV}}_{<k_1+1>_{K_1},k_2}\setminus\{(<k_1+1>_{K_1},k_2)\})\ =\ \{(<k_1+2>_{K_1},k_2),\nonumber \\
&(<k_1+1>_{K_1},<k_2+1>_{K_2}),(<k_1+2>_{K_1},<k_2+1>_{K_2})\} \bigcap \mathcal{T}_3\neq \emptyset. \qquad \ \
\end{align}
Let $\mathcal{S}_3=\{(<k_1+1>_{K_1},k_2)\}\bigcup \mathcal{T}_3$. From \eqref{eq-type-IV} and \eqref{SC1-Z_Uk}, we have $(\mathcal{S}_3\setminus\{(k_1',k_2')\})\bigcap$ $\mathcal{U}^{\text{IV}}_{<k_1+1>_{K_1},k_2} \neq \emptyset$ where $(k_1',k_2')\in \mathcal{S}_3$, then the transmitted coded signal
$X_{\mathcal{S}_3}=\bigoplus_{(k_1',k_2')\in \mathcal{S}_3}$ $W_{d^{\text{IV}}_{\mathcal{U}_{k_1',k_2'}},\mathcal{S}_3\setminus\{(k_1',k_2')\}}$
can be retrieved by user $\mathcal{U}^{\text{IV}}_{<k_1+1>_{K_1},k_2}$, i.e., the packets contained in $X_{\mathcal{S}_3}$ can be retrieved by user $\mathcal{U}^{\text{IV}}_{<k_1+1>_{K_1},k_2}$ through cache nodes $\text{C}_{<k_1+1>_{K_1},k_2}$, $\text{C}_{<k_1+1>_{K_1},<k_2+1>_{K_2}}$, $\text{C}_{<k_1+2>_{K_2},k_2}$ and $\text{C}_{<k_1+2>_{K_1},<k_2+1>_{K_2}}$. Then the server only needs to send the modified coded signal
$X'_{\mathcal{S}_3}=X_{\mathcal{S}_3}-W_{d_{\mathcal{U}^{\text{IV}}_{<k_1+1>_{K_1},k_2}},\mathcal{T}_3}=\bigoplus_{(k'_1,k'_2)\in (\mathcal{S}_3\setminus\{(<k_1+1>_{K_1},k_2)\})}W_{d_{\mathcal{U}^{\text{IV}}_{k'_1,k'_2}},\mathcal{S}_3\setminus\{(k'_1,k'_2)\}}$.
We can check that user $\mathcal{U}^{\text{IV}}_{k_1,k_2}$ can retrieve all the packets in $X'_{\mathcal{S}_3}$ since it can retrieve all packets labeled by $(\mathcal{S}_3\setminus\{(k'_1,k'_2)\})$ where $(k'_1,k'_2)\neq (<k_1+1>_{K_1},k_2)$ from \eqref{SC1-Z_Uk}. We can check that there are
$h^{\text{IV}}_{3}={K_1K_2-3 \choose t}-{K_1K_2-6 \choose t}$
such subsets $\mathcal{T}_3$, so there are $h^{\text{IV}}_{3}$ such transmitted signals in Line 11 of Algorithm \ref{alg-MN} which can be retrieved by user $\mathcal{U}^{\text{IV}}_{k_1,k_2}$.
	
\item Finally, we consider the value $h^{\text{IV}}_4$ by counting the number of $t$-subset $\mathcal{T}_4\in {\mathcal{K} \choose t}$ satisfying the conditions
\begin{align}
\label{cond:IV_T4}
&(k_1,k_2) \notin \mathcal{T}_4,\ \ \ \ \ \ (k_1,<k_2+1>_{K_2}) \notin \mathcal{T}_4,\ \ \ \ \ \  (<k_1+1>_{K_1},k_2) \notin \mathcal{T}_4,\ \ \nonumber \\
&(<k_1+1>_{K_1},\nonumber <k_2+1>_{K_2}) \notin \mathcal{T}_4\ \ \ \ \ \ \text{and}\ \ \ \ \ \ (\mathcal{U}^{\text{IV}}_{<k_1+1>_{K_1},<k_2+1>_{K_2}}\setminus\nonumber \\
&\{(<k_1+1>_{K_1},<k_2+1>_{K_2})\})=\{(<k_1+1>_{K_1},<k_2+2>_{K_2}),\nonumber \\
&(<k_1+2>_{K_1},<k_2+1>_{K_2}),(<k_1+2>_{K_1},<k_2+2>_{K_2})\} \bigcap \mathcal{T}_4\neq \emptyset.
\end{align}
Let $\mathcal{S}_4=\{(<k_1+1>_{K_1},<k_2+1>_{K_2})\}\bigcup \mathcal{T}_4$. From \eqref{eq-type-IV} and \eqref{SC1-Z_Uk}, we have $(\mathcal{S}_4\setminus\{(k_1',k_2')\})\bigcap$ $\mathcal{U}^{\text{IV}}_{<k_1+1>_{K_1},<k_2+1>_{K_2}} \neq \emptyset$ where $(k_1',k_2')\in \mathcal{S}_4$, then the transmitted coded signal
$X_{\mathcal{S}_4}=\bigoplus_{(k_1',k_2')\in \mathcal{S}_4}$ $W_{d^{\text{IV}}_{\mathcal{U}_{k_1',k_2'}},\mathcal{S}_4\setminus\{(k_1',k_2')\}}$
can be retrieve by user $\mathcal{U}^{\text{IV}}_{<k_1+1>_{K_1},<k_2+1>_{K_2}}$, i.e., the packets contained in $X_{\mathcal{S}_4}$ can be retrieve by user $\mathcal{U}^{\text{IV}}_{<k_1+1>_{K_1},<k_2+1>_{K_2}}$ by cache nodes $\text{C}_{<k_1+1>_{K_1},<k_2+1>_{K_2}}$, $\text{C}_{<k_1+1>_{K_1},<k_2+2>_{K_2}}$, $\text{C}_{<k_1+2>_{K_1},<k_2+1>_{K_2}}$ and $\text{C}_{<k_1+2>_{K_1},<k_2+2>_{K_2}}$. Then the server only needs to send the modified coded signal
$X'_{\mathcal{S}_4}=X_{\mathcal{S}_4}-W_{d_{\mathcal{U}^{\text{IV}}_{<k_1+1>_{K_1},<k_2+1>_{K_2}}},\mathcal{T}_4}$ $=\bigoplus_{(k'_1,k'_2)\in (\mathcal{S}_4\setminus\{(<k_1+1>_{K_1},<k_2+1>_{K_2})\})}W_{d_{\mathcal{U}^{\text{IV}}_{k'_1,k'_2}},\mathcal{S}_4\setminus\{(k'_1,k'_2)\}}$.
We can check that user $\mathcal{U}^{\text{IV}}_{k_1,k_2}$ can retrieve all the packets in $X'_{\mathcal{S}_4}$ since it can retrieve all the packets labeled by $(\mathcal{S}_4\setminus\{(k'_1,k'_2)\})$ where $(k'_1,k'_2)\neq (<k_1+1>_{K_1},<k_2+1>_{K_2})$ from \eqref{SC1-Z_Uk}. We can check that there are
$h^{\text{IV}}_{4}={K_1K_2-4 \choose t}-{K_1K_2-7 \choose t}$
such subsets $\mathcal{T}_4$, so there are $h^{\text{IV}}_{4}$ such transmitted signals in Line 11 of Algorithm \ref{alg-MN} which can be retrieved by user $\mathcal{U}^{\text{IV}}_{k_1,k_2}$.
\end{itemize}
	
Finally, by the assumptions
$(k_1,k_2)\in \mathcal{S}_1$, $(k_1,k_2)\notin \mathcal{S}_2$, $(k_1,k_2)\notin \mathcal{S}_3$, $(k_1,k_2)\notin \mathcal{S}_4$, $(k_1,<k_2+1>_{K_2})\in \mathcal{S}_2$, $(k_1,<k_2+1>_{K_2})\notin \mathcal{S}_3$, $(k_1,<k_2+1>_{K_2})\notin \mathcal{S}_4$, $(<k_1+1>_{K_1},k_2)\in \mathcal{S}_3$, and  $(<k_1+1>_{K_1},k_2)\notin \mathcal{S}_4$, we can verify
$\mathcal{S}_1\neq \mathcal{S}_2$, $\mathcal{S}_1\neq \mathcal{S}_3$, $\mathcal{S}_1\neq \mathcal{S}_4$, $\mathcal{S}_2\neq \mathcal{S}_3$, $\mathcal{S}_2\neq \mathcal{S}_4$, and $\mathcal{S}_3\neq \mathcal{S}_4
$. So there are exactly $h^{\text{IV}}=h^{\text{IV}}_{1}+h^{\text{IV}}_{2}+h^{\text{IV}}_{3}+h^{\text{IV}}_{4}$ coded (or modified coded) signals which can be retrieved by user $\mathcal{U}^{\text{IV-1}}_{k_1,k_2}$.

\end{appendices}
\bibliographystyle{IEEEtran}
\bibliography{reference}
\end{document}